\documentclass[a4paper,11pt,hyper]{article}
\usepackage{amsfonts,latexsym,graphicx,epsfig,amssymb,amsmath,mathrsfs,subfig, float, jcappub, wrapfig, ulem}
\pdfoutput=1

\newcommand{\D}{{\rm d}}

\newcommand{\agt}{\hspace{0.3em}\raisebox{0.4ex}{$>$}\hspace{-0.75em}\raisebox{-.7ex}{$\sim$}\hspace{0.3em}}
\newcommand{\alt}{\hspace{0.3em}\raisebox{0.4ex}{$<$}\hspace{-0.75em}\raisebox{-.7ex}{$\sim$}\hspace{0.3em}}

\newcommand{\cA}{{\cal A}}
\newcommand{\Mp}{M_{\text{pl}}}

\title{Resonant magnetogenesis from axions}
\author[a]{Teerthal Patel,}
\author[a]{Hiroyuki Tashiro,}
\author[a,b]{Yuko
	Urakawa}
\affiliation[a]{
	Department of Physics and Astrophysics, Nagoya University, Chikusa,
	Nagoya 464-8602, Japan}
\affiliation[b]{
	Department of Astroparticles and Cosmology, Bielefeld University, Universit{\"a}tsstra{\ss}e 25
	D-33615 Bielefeld 	
}

\abstract{We investigate the generation of seed magnetic field through the Chern-Simons coupling between the U(1) gauge field and an axion field that commences to oscillate at various epoch, depending on the mass scale. We address axions which begin oscillation during inflation, reheating, and also the radiation dominated era after the thermalization of the Universe. We study the resonant generation mechanisms and highlight that a small oscillation time scale with respect to that of the cosmic expansion can lead to an efficient generation of (hyper) magnetic field via resonant generation, even for ${\cal O}(1)$ coupling. In addition, we demonstrate that the generated field can be helical due to the tachyonic amplification phase prior to the onset of oscillation. Furthermore, it is shown that the parametric resonance during reheating can generate a circularly polarized (hyper) magnetic field in a void region with the present amplitude $B_0 =3\times 10^{-15}$Gauss and the coherent length $\lambda_0 = 0.3$pc without being plagued by the backreaction issue. 
}


\begin{document}
	
\maketitle

\section{Introduction}\label{Sec:Introduction}

Magnetic fields have been observed to pervade the Universe on scales ranging from stellar objects to those of galactic clusters. A typical strength on the order of $ \mu \text{G} $ has been observed by Faraday Rotation measurements in galaxies \cite{Wielebinski:2005} and galaxy clusters~\cite{Bonafede:2010xg}. 
Furthermore, there are some reports about the presence of large-scale diffuse synchrotron emission which could be associated with sub-$\mu$G magnetic field in the cosmological filamentary structure~\cite{Vacca:2018rta}.

There are several undetermined aspects of the origin and evolution of such magnetic fields.
There are two classes of the seed magnetic field generation scenarios, an astrophysical origin and a cosmological origin. An interesting aspect that has been considered is that dynamo effects could play an important role in amplifying cosmological seed magnetic field, that existed prior to structure formation in galaxies and galaxy clusters, to their current amplitude (see Refs.~\cite{Davis:1999bt,Pakmor:2013rqa,Barnes:2018ryy}).
Recently, several studies~\cite{Feretti:2012vk,Neronov:1900zz,Tavecchio:2010mk}
suggested the presence of magnetic field in the intergalactic region. They claimed that $ \gamma $-ray observations of TeV blazars placed lower bounds on the magnetic field strength, $B_{\rm Mpc}$, on scales larger than Mpc scales, $B_{\rm Mpc} > 10^{-17}$G, in a void region. If this lower bound in a void is confirmed, it can strongly imply a primordial origin. 

One of the major attempts  in the primordial origin scenarios is to introduce the coupling between the electromagnetic field and a scalar field. Such coupling breaks the conformal invariance of the electromagnetic field, driving a large enhancement of the electromagnetic field.
In this context, the magnetic field generation during inflation has been studied by many authors~(for reviews, see, e.g.,~Refs.~\cite{Martin:2007ue, Durrer:2010mq, Caprini:2014mja, Fujita:2015iga, Ratra:1991bn, Ferreira:2013sqa}).
It has been suggested that an axion or axion-like field may become a prominent source of the cosmological magnetic field~\cite{Garretson:1992vt, Finelli:2000sh} (also see, e.g., Refs.~\cite{Sikivie:2006ni, Marsh:2015xka} for a review about axions in cosmology). An axion $\phi$ is typically coupled with the electromagnetic field through the Chern-Simons coupling $\phi \, \epsilon^{\mu \nu \rho \sigma} F_{\mu\nu}{F}_{\rho \sigma}$, where ${F}_{\mu \nu}$ is the electromagnetic field strength tensor. Since the Chern-Simons coupling violates the parity symmetry, the electromagnetic field generated through this coupling can be circularly polarized, having a non-zero helicity. 
The successive evolution of a helical magnetic field in cosmological plasma is qualitatively different from that of a non-helical one (for generation of non-helical magnetic field, see e.g., Refs.~\cite{Durrer:2013pga,Subramanian:2015lua}). 
When the helicity is conserved in time, a fraction of the magnetic field energy on small scales
is transferred to large scales, i.e., the inverse cascade takes place~\cite{Banerjee:2004df}. As a result, the coherent scale of the helical magnetic field can become much larger than the one at generation. 
Another interesting aspect of primordial helical magnetic fields was discussed in Ref.~\cite{Giovannini:1997eg} and more recently in Refs.~\cite{Anber:2015yca, Fujita:2016igl, Kamada:2016eeb, Kamada:2016cnb}, where baryogenesis sourced by helical magnetic fields was discussed (see also Refs.~\cite{Domcke:2018eki, Domcke:2019mnd}).
Meanwhile, recently it was reported that the spatial distribution of the observed $\gamma$-ray background suggests the existence of helical cosmological magnetic field~\cite{Tashiro:2013ita}.

So far, magnetogenesis from an axion (-like field) $\phi$ has been largely explored, assuming that $\phi$ has driven the inflationary expansion in the early universe, for example ~\cite{Durrer:2010mq,Caprini:2014mja, Fujita:2015iga, Turner:1987bw, Adshead:2016iae}. In Ref.~\cite{Durrer:2010mq}, it was demonstrated that the amplification of the U(1) gauge field during the slow-roll regime of $\phi$ via the tachyonic instability is insufficient in satisfying the blazar bounds. Ref.~\cite{Fujita:2015iga} showed that the tachyonic instability of $\phi$ just before the end of inflation can further enhance the seed magnetic field. A successful energy transfer from $\phi$ to the gauge field requires, either a large growth rate or long-lasting growth span. The former can be realized by postulating a large coupling constant for the Chern-Simons coupling. Meanwhile, in an expanding universe, the redshift due to the cosmic expansion tends to disturb a sustained parametric resonance. Recently, Ref.~\cite{Kitajima:2018zco} revealed that this is not always the case, in particular when $\phi$ starts to oscillate at a time much later than when the time scale of the cosmic expansion becomes comparable to that of the oscillation. Such a delayed onset of the oscillation can be realized, when the curvature of the potential for the oscillation regime is substantially larger than the one for the slow-roll regime. While the onset of the oscillation is determined by the latter, once $\phi$ starts to oscillate, the time scale of the oscillation is predominantly determined by the former. This provides an alternative scenario for the successful energy transfer, which does not require a large coupling with the inflaton. In Ref.~\cite{Kitajima:2018zco}, it was shown that parametric resonance of the axion's self interaction can continue without being disturbed by cosmic expansion, leading, in turn, to a copious production of gravitational waves, when the onset of the oscillation delays. In this paper, we will exhibit that it is also the case for the U(1) gauge field which is coupled with the oscillating axion, via the Chern-Simons coupling.  

Not only being a candidate of the inflaton and dark matter, axions provide a unique window to explore a theoretical prediction of string theory. It is known that string theory predicts a copious presence of axions whose mass spectrum is logarithmically flat. The Universe which is filled with such axions is called string axiverse~\cite{Arvanitaki:2009fg}. Along the same line, in this paper, we investigate the possibility of a magnetogenesis through axions which start to oscillate at various epochs, including after the completion of the reheating. A major obstacle to generate the magnetic field in the plasma filled Universe is the presence of the large electrical conductivity, which introduces a friction whose time scale is much shorter than that of the cosmic expansion. In this paper, we address whether the parametric resonance caused by the rapid coherent oscillation which follows the delayed onset of the oscillation can generate the magnetic field by overcoming the friction due to the conductivity, or not.

When the onset of the oscillation significantly delays, the parametric resonance continues until the backreaction of the gauge field production becomes important. At the non-linear regime, the two polarization modes, which had evolved independently in the linear regime, start to interact with each other, as explored in Ref.~\cite{Adshead:2015pva}. This washes out the helicity generated through the Chern-Simons coupling in the linear regime. Therefore, in this paper, focusing on the models where the onset of the oscillation mildly delays and the parametric resonance terminates due to the cosmic expansion before the saturation, we estimate the maximum amplitude of the present magnetic field. 

This paper is organized as follows. In Sec.~\ref{Sec:Preliminaries}, we first discuss the general formulation governing the co-evolution of the U(1) gauge field and the axion. Then, we summarize the various setups of the magnetogenesis which are addressed in this paper. In Sec.~\ref{SSec:BR}, we derive a general formula for the possible amplitude of the magnetic field at generation without being plagued by the backreaction problem. In Sec.~\ref{Sec:gauge field generation during inflation}, we discuss the gauge field production during inflation. As a specific example, we consider axion monodromy inflation. In Sec.~\ref{Sec:production during axion reheating}, we discuss the gauge field production during reheating. In Sec.~\ref{Sec:gauge field generation during inflation} and Sec.~\ref{Sec:production during axion reheating}, we assume that the axion has driven the inflationary expansion as the dominant source. In Sec.~\ref{Sec:spectator production}, we consider the gauge field production by a spectator axion both during inflation and reheating. In Sec.~\ref{Sec:production during axion reheating} and Sec.~\ref{Sec:spectator production}, we discuss the gauge field production through parametric resonance, when the onset of the oscillation delays. In Sec.~\ref{Sec:post inflationary gen}, we address the resonant production of the gauge field in conducting plasma. In Sec.~\ref{Sec:present}, we provide the order estimation of the possible magnetic field amplitude at present for the scenarios discussed in the paper.

\section{Preliminaries}\label{Sec:Preliminaries}
In this paper, we consider a U(1) gauge field $A^\mu$ and an axion (or an axion-like field) $\phi$ with the Lagrangian density given by
\begin{align}
\label{Lagrangian axion inflaton coupled to EM}
& {\cal L} = - \frac{1}{2} \partial_\mu \phi \partial^\mu \phi -
V(\phi) - \frac{1}{4} F_{\mu \nu} F^{\mu \nu} - \frac{\alpha}{4} \frac{\phi}{f}
F_{\mu \nu} \tilde{F}^{\mu \nu}\,,
\end{align}
where $ \alpha $ is the dimensionless coupling parameter and $ f $ is the decay constant of the axion. Since the potential $V(\phi)$ preserves the $Z_2$ symmetry, we can choose $\alpha$ to be positive, without a loss of generality. Here $ F_{\mu\nu} $ is the field strength of the U(1) gauge field defined as $ F_{\mu\nu} = \partial_\mu\mathcal{A}_\nu - \partial_\nu\mathcal{A}_\mu $ and the dual $ \tilde{F}_{\mu\nu} $ is defined as 
\begin{equation}
	\label{F Dual}
	\tilde{F}^{\mu\nu} = \frac{1}{2\sqrt{-g}} \, \epsilon^{\mu\nu\rho\sigma}F_{\rho\sigma}\,,
\end{equation}
where $ \epsilon^{\mu\nu\rho\sigma} $ is the rank-four Levi-Civita tensor and we adopt the convention $ \epsilon^{0ijk} = \epsilon^{ijk}$. The temporal (or spatial) variation of $\phi$ breaks the conformal symmetry in the gauge field sector through the Chern-Simons coupling, leading to a possible amplification of the U(1) gauge field.

\subsection{Gauge field production in different epoch}\label{SSec:gauge field generation at various epoch}
An axion commences to oscillate when the mass roughly becomes comparable to the Hubble parameter $H$. The axion plays the role of dark matter when it oscillates coherently around the potential minimum. If our Universe is filled with multiple axions in a wide mass range, as was argued in the context of string axiverse~\cite{Arvanitaki:2009fg}, the axions start to oscillate at different epochs in the history of the Universe. Along this line, in this paper, we investigate a generation of the U(1) gauge field through the Chern-Simons coupling with an axion (or an axion-like field) which commences oscillation at various epochs. In particular, we consider the following four different setups:
\begin{enumerate}
    \item Gauge field production during inflation
    \begin{enumerate}
        \item Axion $=$ Inflaton  ($\to$ Section ~\ref{Sec:gauge field generation during inflation})
        \item Axion $\neq$ Inflaton   ($\to$ Section ~\ref{Sec:spectator production})
    \end{enumerate}
    
    \item Gauge field production during reheating   ($\to$ Section ~\ref{Sec:production during axion reheating})
    
    \item Gauge field production during radiation or matter domination  ($\to$ Section~\ref{Sec:post inflationary gen})
\end{enumerate}
In this paper, considering the case where the symmetry breaking happened before the end of inflation, we assume that the initial distribution of the axion is almost homogeneous. As will be discussed later, the generation of the gauge field proceeds quite differently before and after the onset of the oscillation. 
In particular, the coherent oscillation of the axion has been exploited as the driving source for the copious production of particles via parametric resonance~\cite{Kofman:1997yn}. 

When the axion is the dominant component of the Universe, the time evolution of the axion has to satisfactorily explain the expansion history of the Universe both during and post inflation. When the axion is the inflaton, the onset of the oscillation corresponds to the beginning of the reheating process. Meanwhile, as far as the axion is subdominant, the axion dynamics is not strictly limited by the expansion history. For the cases 1-(b), 2 and 3, the axion is not (necessarily) the dominant component of the Universe at the onset of oscillation. 

The gauge field production discussed in this paper serves a mechanism for magnetogenesis when the axion is coupled with the U(1) gauge field in the standard model sector. Before and after the electro-weak phase transition, we can identify $A^{\mu}$ as the U(1)$_Y$ gauge field and the electromagnetic field, respectively. It was argued that the generated U(1)$_Y$ gauge field is efficiently transferred into the electromagnetic field after the electro-weak phase transition~\cite{Dimopoulos:2001wx}. Therefore, in the following, we estimate the amplitude of the magnetic field without distinguishing these two U(1) gauge fields. 

An axion acquires a potential through a non-perturbative effect such as an instanton effect. It is widely known that when the dilute instanton gas approximation holds, the acquired potential of the axion is given by the cosine form as
\begin{equation}
V (\phi) = (mf)^2 \left[ 1- \cos{\left(\frac{\phi}{f}\right)} \right]\,.  \label{Exp:Vcos}
\end{equation}
However, as was first pointed out in Refs.~\cite{Witten:1979vv, Witten:1980sp}, this form does not necessarily hold when this approximation is broken. Recently, it was shown that when the axion is strongly coupled with an SU($N$) gauge field, the axion's potential is given by~\cite{Nomura:2017ehb}
\begin{equation}
V(\phi) = \Lambda^4 \left[ 1 - \frac{1}{(1+ c \phi^2)^p} \right]\,, \label{Exp:Vpn}
\end{equation}
in the large $N$ limit (see also Ref.~\cite{Yonekura:2014oja}). Here, $c$ and $p$ are positive constant parameters. In this paper, we consider this general class of potentials. Both Eqs.~(\ref{Exp:Vcos}) and (\ref{Exp:Vpn}) have a shallower region than the quadratic potential. The validity of the dilute instanton gas approximation and the potential form of the axion have been better investigated for the QCD axion which interacts with the thermalized QCD fields~(see e.g., \cite{Berkowitz:2015aua, Borsanyi:2016ksw}). In Refs.~\cite{Soda:2017dsu, Kitajima:2018zco}, it has been shown that, when a spectator axion was initially located in such a shallow potential region, it starts to oscillate much later than $H\sim m$, where $m$ is defined by the potential curvature at the bottom of the potential.
As a result, the time scale of the cosmic expansion becomes much longer than that of the  oscillation. Since the cosmic expansion no longer disturbs the resonant growth, a strong resonant amplification can be realized.
Refs.~\cite{Soda:2017dsu, Fukunaga} have studied the parametric resonance driven by the self-interaction of the axion. In this paper, we will demonstrate that the parametric resonance, driven by the Chern-Simons interaction, is also extremely efficient for the delayed oscillation case, $H_{\rm osc}/m \ll 1$. 

\subsection{Basic formulae}\label{SSec:Basic formulae}
In this section, we summarize the basic formulae which will be used in this paper. Considering an observer with 4-velocity $ u^\mu $, we decompose the U(1) gauge field $A^{\mu}$ as 
\begin{equation}
	\label{Def:EB}
	E_\mu \equiv F_{\alpha\mu}u^\alpha,\qquad B_\mu \equiv  \tilde{F}_{\mu\alpha}u^{\alpha}\,.
\end{equation}
For the electromagnetic field, Eq.~(\ref{Def:EB}) provides a covariant definition of the electric and magnetic fields. 
In this paper, we adopt the Coulomb gauge:
\begin{equation}
     A^0 = \partial_iA^i = 0\,.
\end{equation}
For a comoving observer in the FLRW (Friedmann-Lema\^{i}tre-Robertson–Walker) metric, whose 4-velocity is given by $ u^\mu = a^{-1}(1,\mathbf{0}) $, the $ E $ and $ B $ field strengths can be expressed as
\begin{equation}
	\label{E and B field strength in terms of the vector potential in FLRW adn Coulomb gauge}
	E_\mu = \left( 0 , -\frac{1}{a}A'_i \right), \qquad B_\mu = \left( 0, \frac{1}{a}\epsilon_{ijk}\partial_jA_k \right)\,,
\end{equation}
where  the prime denotes derivative w.r.t conformal time $ \eta $. 

Using the annihilation and creation operators $ b^{(\pm)}_{\mathbf{k}} $ and $b^{(\pm)\dagger}_{\mathbf{k}} $, which satisfy the commutation relation, 
\begin{equation}
	    [b^{(h)}_{\mathbf{k}}, b^{(h')\dagger}_{\mathbf{k'}}] = (2\pi)^3\delta(\mathbf{k} - \mathbf{k}')\delta^{h,\, h'}\,, \qquad \quad (h, h' = \pm)\,,
\end{equation}
we can quantize the gauge field $A_i$ as
\begin{equation}
	\label{Fourier expansion of vector potential in terms of polarizaton vectors}
	A_i(\eta,\,\mathbf{x}) = \sum_{h=\pm}^{}{ \int{ \frac{\text{d}^3 {\mathbf k} }{(2\pi)^3} e^{ i\mathbf{k}\cdot\mathbf{x} } e^{(h)}_{i} (\mathbf{\hat{k}})[ b^{(h)}_{\mathbf{k}}\mathcal{A}_h(\eta,\, k) + {b^{(h)\dagger}_{-\mathbf{k}}}\mathcal{A}_h^* (\eta,\, k)]    }}\,.
\end{equation}
	Here, $\hat{\mathbf{k}}$ is the unit vector defined as $\hat{\mathbf{k}} \equiv \mathbf{k}/k$ and $e^{(h)}_i(\mathbf{\hat{k}})$ with $h= \pm$ is the polarization vector which satisfies $\hat{{k}}^i e^{(h)}_i(\mathbf{\hat{k}})=0$ and $ i\epsilon_{ijl} \hat{k}_j e_l^{(h)}(\mathbf{\hat{k}}) = h e^{(h)}_i(\mathbf{\hat{k}}) $. 

Taking the derivative of the Lagrangian density given in Eq.~(\ref{Lagrangian axion inflaton coupled to EM}), the equations of motion for the homogeneous mode of the axion $ \phi $ and the Fourier modes of $ \mathcal{A}_h $ in the FLRW background are given by
\begin{align}
& \phi'' + 2\mathcal{H}\phi' + a^2V_{,\phi} = a^2  \frac{\alpha}{f}\langle \mathbf{E}\cdot\mathbf{B}\rangle\,, \label{Phi EOM} \\
&\mathcal{A}_h''(\eta,\, k)+\left(k^2 - h\frac{\alpha}{f}\frac{\phi'}{\mathcal{H}}k\mathcal{H}\right)\mathcal{A}_h(\eta,\, k) = 0\,. \label{Vector field fourier mode EOM }
\end{align}
The coupling between the axion and the gauge field is characterized by
\begin{equation}
    \xi \equiv \frac{\alpha}{f} \frac{\phi'}{\mathcal{H}}\,. \label{Def:xi}
\end{equation}
The term in the right hand side of Eq.~(\ref{Phi EOM}), which describes the backreaction of a gauge field production on $\phi$, is given by
\begin{equation}
\langle \mathbf{E} \cdot \mathbf{B}\rangle = -  \frac{1}{8\pi^2}\!\int{ \text{d}\ln k \left(\frac{k}{a}\right)^{\!4}}\left[ \frac{1}{k}\frac{\text{d}}{\text{d}\eta}\left( |\sqrt{2k}\mathcal{A}_+|^2\right) - \frac{1}{k}\frac{\text{d}}{\text{d}\eta}\left( |\sqrt{2k}\mathcal{A}_-|^2\right) \right] \,.\label{Exp:EdotB}
\end{equation}
Here, we expressed the gauge field in the combination $\sqrt{2 k } {\cal A}_h$, since we will impose the WKB initial condition, where the amplitude of $\sqrt{2 k } {\cal A}_h$ is initially set to $1$. Reflecting the fact that $\langle\mathbf{E} \cdot \mathbf{B}\rangle$ violates the parity symmetry, the right-handed and left-handed polarization modes contribute to it with opposite signatures.

When the axion $\phi$ monotonically evolves in time, taking a positive or negative value for $\phi'$, the Fourier modes with $k/{(aH)} < |\xi|$ for one of the two polarizations $h= \pm$ exponentially grows due to the tachyonic instability. Once $\phi$ starts to oscillate, the tachyonic instability can take place for both polarization modes.

In terms of the mode function ${\cal A}_h(\eta,\, k)$, the energy density for the gauge field is given by $\rho_{\rm em} = \rho_{\rm E} + \rho_{\rm B}$ with
\begin{align}
	& \rho_{\rm E}(\eta) \equiv  \frac{1}{4 \pi^2} \int \frac{\text{d} k}{k} \left( \frac{k}{a} \right)^4
   \sum_{h= \pm} \Bigl| \frac{1}{k} \frac{\text{d}  \sqrt{2 k}{\cal A}_h}{ \text{d}\eta} \Bigr|^2\,, \label{Exp:RhoE}\\
	& \rho_{\rm B}(\eta) \equiv \frac{1}{4 \pi^2} \int \frac{\text{d} k}{k} \left( \frac{k}{a} \right)^4 \sum_{h= \pm} \Bigl| \sqrt{2 k} {\cal A}_h \Bigr|^2\,. \label{Exp:RhoB}
\end{align}
For the electromagnetic field, $\rho_{\rm E}$ and $\rho_{\rm B}$ correspond to the energy densities of the electric and magnetic fields, respectively. Using ${\cal A}_h$, the helicity density for the gauge field is given by
\begin{align}
	& \langle \mathfrak{H} \rangle 
	= \int \frac{\text{d}^3\mathbf{k}}{(2\pi)^3}k(|\mathcal{A}_+|^2-|\mathcal{A}_-|^2)\label{Exp:Helicity}\,.
\end{align}	
The Chern-Simons coupling breaks the parity symmetry. Since it can enhance either the right-handed or left-handed polarization, the helicity density can increase or decrease in time. In this paper, we assume that other than this coupling between the axion and the U(1) gauge field, there is no other process which induces a time variation in the helicity density.

Using Eq.~(\ref{Exp:RhoB}), we define the amplitude of the magnetic field (or the hypermagnetic field) with the wavenumber $k$ as
\begin{equation}
B_h(\eta, k) \equiv \sqrt{2\frac{\text{d}\ln \rho_B}{\text{d}\ln k}} =  \frac{1}{\sqrt{2} \pi} \left(\frac{k}{a} \right)^2  \sqrt{2 k} \Bigl|{\cal A}_h (\eta,\, k)\Bigr| \,,  \label{Def:B}
\end{equation}
for the two polarization modes $h = \pm$.

\subsection{Estimation of the backreaction}  \label{SSec:BR}
Next, let us discuss an order estimation of the possible amplitude of the (hyper) 
magnetic field by imposing 
\begin{equation}
    \rho_{\rm E}\,, \rho_{\rm B} \alt  \rho_\phi \,,  \label{Exp:BRrho}
\end{equation}
and
\begin{equation}
    \left| \frac{\alpha}{f}\langle \mathbf{E}\cdot\mathbf{B}\rangle \right| \alt \left| V_{,\, \phi} \right|\,. \label{Exp:BReom}
\end{equation}
{When the axion is the only source of the gauge field production, the condition (\ref{Exp:BRrho}) should be generically fulfilled. Meanwhile, a violation of the condition (\ref{Exp:BReom}) is not immediately problematic, particularly when the axion is not the dominant component of the Universe. However, the resonant production of the gauge field, which turns out to be the most efficient production mechanism among those studied in this paper, typically terminates, when the inequality in Eq.~(\ref{Exp:BReom}) is saturated. }
Since $V_{,\, \phi}$ oscillates and passes through 0, e.g., during reheating, the right hand side of Eq.~(\ref{Exp:BReom}) should be understood as the amplitude of the oscillating $V_{,\, \phi}$.

Using Eqs.~(\ref{Exp:RhoE}) and (\ref{Exp:RhoB}), under the assumption that the spectrum of $\cA_h$ has a peak around $k = k_{m, {\rm gen}}$ at the generation, the first condition (\ref{Exp:BRrho}) can be rewritten in	
\begin{align}
   & \Delta (\ln k) \left( \frac{k_{m, {\rm gen}}}{a_{{\rm gen}} H_{{\rm gen}}} \right)^4 \frac{1}{F_{\rm gen}}  \left( \frac{H_{\rm gen}}{\Mp} \right)^2 \Bigl| \frac{1}{k} \frac{\text{d}  \sqrt{2 k} {\cal A}_h}{ \text{d}\eta} \Bigr|^2 \alt 1\,, \label{Exp:RhoE2} \\
   & \Delta (\ln k) \left( \frac{k_{m, {\rm gen}}}{a_{{\rm gen}} H_{{\rm gen}}} \right)^4 \frac{1}{F_{\rm gen}}  \left( \frac{H_{\rm gen}}{\Mp} \right)^2 \Bigl| \sqrt{2 k} {\cal A}_h \Bigr|^2 \alt 1 \,, \label{Exp:RhoB2}
\end{align}	
where we introduced the fraction of the axion energy density to the total energy density at the generation as
\begin{align}
    F_{{\rm gen}} \equiv \frac{\rho_{\phi,\, {\rm gen}}}{\rho_{{\rm tot},\, {\rm gen}}}\,.
\end{align}
Here, $\Delta \ln k $ measures the width of the peak in the spectrum of $\cA_h$. When the axion $\phi$ is the dominant component of the Universe, $F_{{\rm gen}}$ can be approximated as 1. Here and hereafter, we express a quantity evaluated at the generation of the gauge field with the subscription ${\rm gen}$. Both of the two polarization modes should satisfy Eqs.~(\ref{Exp:RhoE2}) and (\ref{Exp:RhoB2}). Meanwhile, using Eq.~(\ref{Exp:EdotB}), the second condition (\ref{Exp:BReom}) yields
\begin{align}
     \Delta (\ln k) \left( \frac{k_{m, {\rm gen}}}{a_{{\rm gen}} H_{\rm gen}} \right)^4 \alpha \frac{\Mp}{f} \left| \frac{1}{k}\frac{\text{d}}{\text{d}\eta} |\sqrt{2k}\mathcal{A}_h|^2 \right|  \alt \frac{\Mp | V_{, \phi} |} {H_{\rm gen}^4} \,. \label{Exp:BReom2}
\end{align}
These conditions give the upper bounds on the amplitudes of (the growing mode of) $\cA_h$ and its time derivative. 

Combining these three conditions (\ref{Exp:RhoE2}), (\ref{Exp:RhoB2}), and (\ref{Exp:BReom2}) and using Eq.~(\ref{Def:B}), we obtain the upper bound on the (hyper) magnetic field at the peak wavenumber as
\begin{align}
    B_h (\eta_{{\rm gen}}, k_{m, {\rm gen}}) \alt \frac{H_{{\rm gen}} \Mp}{\sqrt{\Delta \ln k}} \times {\rm min} \left[ \sqrt{F_{\rm gen}},\,\, \frac{\sqrt{F_{\rm gen}}}{D},\,\,\frac{1}{\sqrt{\alpha D}} \Bigl(\frac{f}{\Mp} \Bigr)^{1/2} \frac{|V_{, \phi}|^{1/2}}{H_{\rm gen} \Mp^{1/2}} \right] \,. \label{Cond:B}
\end{align}
Here, we introduced
\begin{align}
    D(\eta,\, k) \equiv \frac{1}{|\mathcal{A}_h|} \Bigl| \frac{1}{k} \frac{\text{d} {\cal A}_h}{ \text{d}\eta} \Bigr|\,,
\end{align}
and we approximated $|d | \cA_h|^2/d {(k \eta)}| \alt D |\mathcal{A}_h|^2$. Eq.~(\ref{Cond:B}) states that, when the width of the peak is narrower, i.e., $\Delta \ln k \ll 1$, a larger peak amplitude of $B$ is allowed without contradicting the conditions (\ref{Exp:BRrho}) and (\ref{Exp:BReom}). 

In this work, we discuss the magnetic field amplitude that can be generated without violating these conditions. In all the numerical analyses, we assume that the generation of the gauge field does not significantly modify the cosmic expansion and the evolution of the background $\phi$. 
This is simply accomplished by requesting that the conditions (\ref{Exp:BRrho}) and (\ref{Exp:BReom}) are satisfied by a substantial margin. 
As is clear from \eqref{Cond:B}, a larger energy density of the source, $\rho_{\phi, {\rm osc}} \propto F_{\rm osc}H^2_{\rm osc}$, or generating the gauge field with a smaller coupling constant $\alpha$ enhances the possible amplitude of $B_h$, thus relaxing the restriction. 
In the numerical analyses, we further require that the inhomogeneous contribution of the axion in the equation of motion for ${\cal A}$ is suppressed. In this paper, we assume that the axion coherently oscillates at least within each Hubble patch. This assumption should be more carefully examined especially after the non-linear structure formation of the Universe sets in (see, e.g., Ref.~\cite{Guth:2014hsa}).    

\section{Gauge field production during inflation: Axion = Inflaton}\label{Sec:gauge field generation during inflation}
In this section, we consider the generation of the gauge field by the axion which plays the role of inflaton. As seen from Eq.~(\ref{Vector field fourier mode EOM }), the coupling between the axion and the gauge field is proportional to $\xi$, defined in Eq.~(\ref{Def:xi}). Using the slow-roll parameter,
\begin{equation}
    \varepsilon \equiv \frac{1}{2 \Mp^2} \left( \frac{\phi'}{{\cal H}} \right)^2\,,
\end{equation}
we can express the amplitude of the parameter $\xi$ as 
\begin{align}
    |\xi| =\sqrt{2 \varepsilon} \alpha \frac{\Mp}{f}\,.  \label{Exp:xi}
\end{align}
Therefore, in slow-roll inflation with $\varepsilon \ll 1$, the amplitude of $|\xi|$ is bounded as $|\xi| \ll \alpha(\Mp/f)$ (in the canonically normalized frame). In fact, this is one obstacle in amplifying the gauge field efficiently during slow-roll inflation.

\subsection{Gauge field production in monodromy inflation}\label{SSec:generation in monodromy}
During inflation, we determine the initial condition for ${\cal A}_h$ in the limit $k/{\cal H} \gg 1$, requiring that ${\cal A}_h$ should approach to the WKB solution, given by
\begin{equation}
\label{WKB initial conditions}
\mathcal{A}_h (\eta,\, k) \rightarrow \frac{1}{\sqrt{2k}}e^{-ik\eta}\,.
\end{equation}
When the slow-roll parameter $\varepsilon$ almost stays constant in time, we can solve the mode equation for $\cA_h$ analytically
Ref.~\cite{Stegun}. In this case, the asymptotic value for the growing mode of $\cA$ in the limit $- k \eta \to 0$ can be written as~\cite{Anber:2006xt}
\begin{equation}
\label{Asymptotic growning solution}
\mathcal{A}(\eta,\, k) \xrightarrow{|k\eta| \ll 1 }\frac{1}{\sqrt{2k}}\frac{e^{\pi|\xi|}}{\sqrt{2\pi|\xi|}}\,.
\end{equation}

During inflation, $\varepsilon$ varies in time, increasing towards the end of inflation. When $\xi$ is time dependent, in general, we need numerical analysis to solve for the evolution of $\cA_h$. As a concrete example, we consider the generation of magnetic field in axion monodromy inflation {~\cite{Silverstein:2008sg}}, where the potential is given by 
\begin{equation}
\label{Monodromy Potential}
V(\phi) = \mu^{4-p}\phi^p + \Lambda^4 \cos{\frac{\phi}{f}}\,.
\end{equation}
The power-law term in $V$ drives the inflationary expansion of the Universe. In addition, there is an oscillatory correction term characterized by the amplitude $\Lambda^4$ and the decay constant $f$. In the following, we choose $f$ so that the frequency of the oscillatory contribution in the background $\phi$ satisfies $H < \omega < \Mp$. In the following, we set $p$ to $p=2$, focusing on around the end of inflation, where the coupling between $\phi$ and the gauge field can be the most effective during inflation.

\begin{figure}
	\centering
	\includegraphics[width=1.\linewidth]{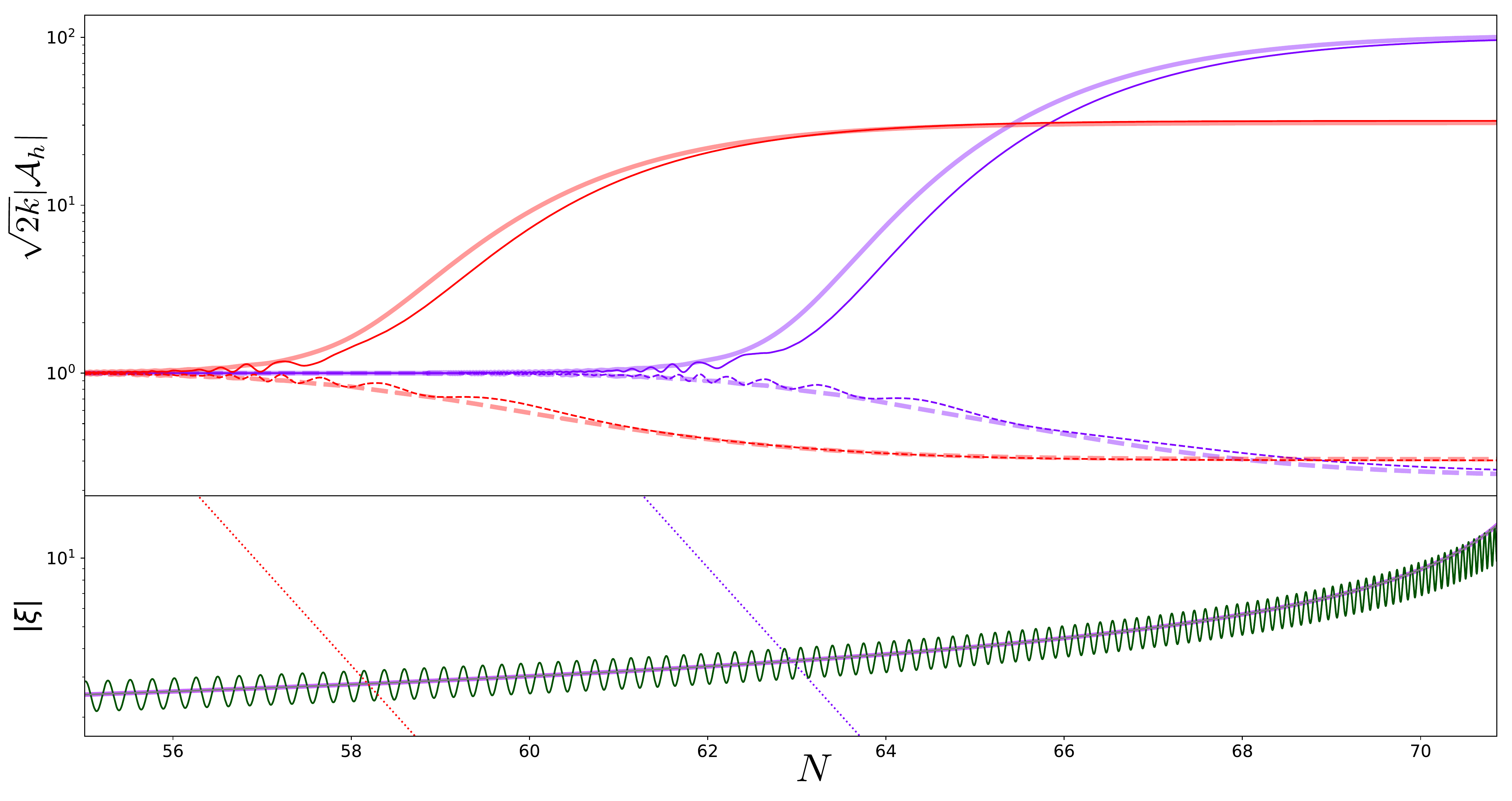}
	\caption{The lower and upper panels show the evolution of $\xi$ and ${\cal A}_h$, respectively, for axion monodromy inflation, where the scalar potential is given by Eq.~(\ref{Monodromy Potential}), with the parameters $\alpha=0.1$, $f/\Mp = 0.01$, and $p=2$. The time coordinate in horizontal axis is the $e$-folding number $N$. The red and violet sets of solid ($h=-$) and dashed ($h=+$) lines in the upper panel correspond to modes with $k/(a_f H_f) = \exp(-12)$ and $\exp(-7)$, respectively, i.e. the modes that crossed the Hubble scale roughly 12 and 7 $e$-foldings before the end of inflation and the corresponding evolution of $k/(aH)$ are represented by the dotted lines in the lower panel. The thin dark and thick light curves show the evolution of ${\cal{A}}_h$ for $\Lambda^4/(\mu \Mp)^2 = 0.2$ and $0$, respectively . In the lower panel, the green and violet curves correspond to the evolution of $|\xi|$ with $\Lambda^4/(\mu \Mp)^2 = 0.2$ and $0$, respectively. 
	}	\label{fig:ev_monodoromy}
\end{figure}

Figure \ref{fig:ev_monodoromy} shows the time evolution of $|\xi|$ and $\cA_{\pm}$, when the scalar potential of the axion is given by Eq.~(\ref{Monodromy Potential}) without and with the oscillatory contribution, i.e., $\Lambda = 0$ and $\Lambda^4/(\mu \Mp)^2 = 0.2$, respectively. 
In the numerical computation, we evaluated the co-evolution of the axion and gauge field in terms of the time variable $e$-folding number $N$, starting from $k/a > \omega$~(where $\omega$ is the frequency of the background oscillation) for several wavenumbers $k$ until $\varepsilon$ reaches 1. Similar to the spectra of the curvature perturbation in monodromy inflation~\cite{Flauger:2009ab, Flauger:2010ja}, the oscillation in the scalar potential leads to oscillatary features in both the axion and the gauge field. In the upper panel of Fig.\ref{fig:ev_monodoromy}, we show the time evolution of $\cA_{\pm}$ for two wavenumbers. In the lower panel, the dotted lines show the time evolution of $k/(aH)$ for the two corresponding wavenumbers. The point at which the solid lines and dotted lines intersect are the moments where $k/(aH) \simeq |\xi|$ for these two wavenumbers. One of the two polarization modes starts to grow after crossing $k/(aH) \simeq |\xi|$. We find that even in the presence of oscillatory contribution in $V$, the asymptotic value of the growing mode for each wavenumber $k$ roughly agrees with the one for the slow-roll evolution, given in \eqref{Asymptotic growning solution}, when we use the value of $|\xi|$ at the intersection, $k/(aH) \simeq |\xi|$. The slow-roll parameter $\varepsilon$ reaches ${\cal O} (1)$ around $N\simeq70$, terminating the inflationary expansion.
When the background trajectory for $\Lambda=0$ preserves the slow-roll approximation and the amplitude of the oscillatory contribution in $V$ is perturbative, we can analytically solve the time evolution of ${\cal A}$. However, here, to capture the ramp up of $\varepsilon$ towards the end of inflation, we solved the time evolution of the axion and the gauge field numerically, taking into account the cosmic expansion consistently.

\begin{figure} 
	\centering
	\includegraphics[width=1.\linewidth]{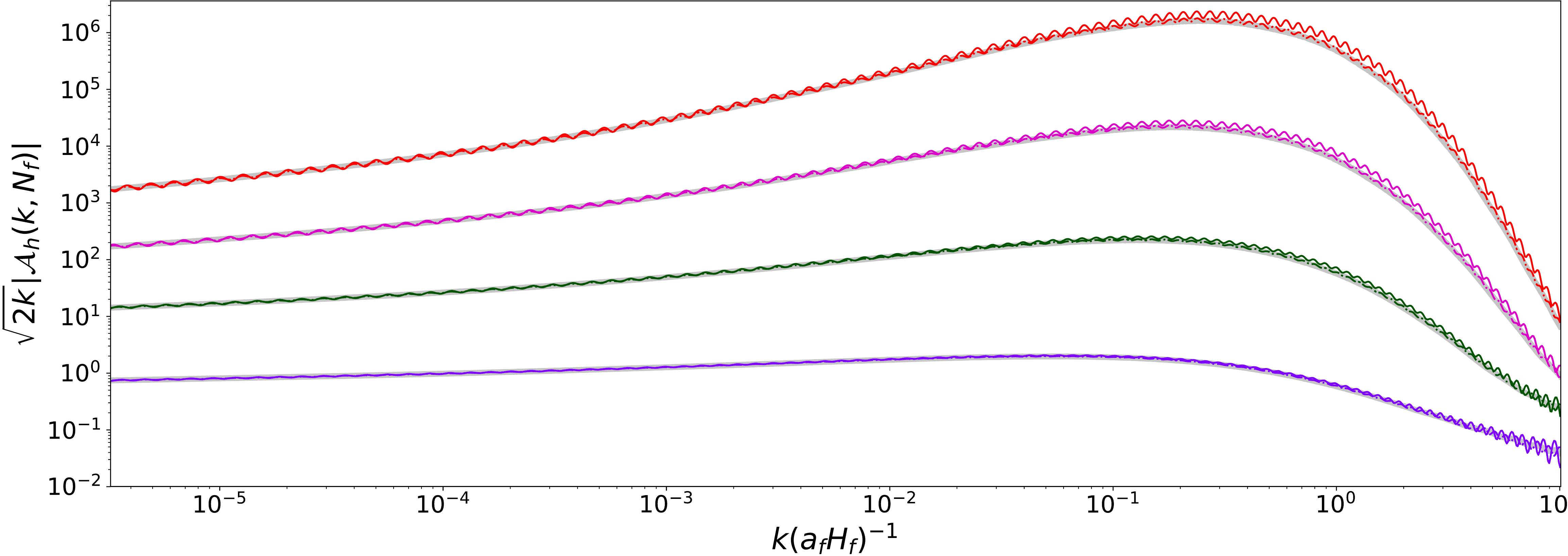}
	\caption{
	The violet, green, magenta and red curves correspond to the spectra of the growing mode($\mathcal{A}_-$) for $\alpha = 0.02, 0.08, 0.14, 0.2$, respectively and the decay constant here is set to $ f/\Mp = 0.01 $. The grey bands, dashed-dotted and solid curves represent the spectra  for oscillatory parameters $ \Lambda^4/(\mu \Mp)^2 = 0,0.1\text{ and }0.2 $, respectively.}  
	\label{fig:monodromy spectra}
\end{figure}

Figure~\ref{fig:monodromy spectra} shows the spectrum with $ \ln k/(a_f H_f) \in [-12,3] $ evaluated at different times. As can be seen, a larger value of $ \alpha $ results into a larger peak amplitude. In the plot, only the dominant polarization mode is shown but it was confirmed that the other mode is damped away. We chose the oscillation amplitude as $ \Lambda^4/(\mu \Mp)^2 =0,\, 0.1,\, 0.2$, for which the slow-roll parameter $\varepsilon$ remains smaller than 1 for a sufficient number of $e$-folds. As shown in Fig.~\ref{fig:monodromy spectra}, the oscillatory contribution does not lead to a significant amplification. This is because, as we argued, the asymptotic value of the growing mode is roughly given by Eq.~\eqref{Asymptotic growning solution} even in the presence of the oscillatory contribution. As one can imagine from the expression 
Eq.~\eqref{Asymptotic growning solution}, the final amplitude becomes larger for a larger $|\xi|$. For $\Lambda=0$, since the amplitude of $|\xi|$ increases towards the end of inflation, the amplitude of the asymptotic value is larger for $k$, which crosses $k/(aH) \simeq |\xi|$ later. The amplitude starts to decrease around $k/(a_f H_f) \simeq \exp(-2) $, because the wavenumbers $k/(a_f H_f) \agt \exp(-2)$ do not have enough time to reach the asymptotic value up until the time when $\varepsilon$ becomes $\mathcal{O}(1)$. As a result, the spectrum creates a peak at $k/(a_f H_f) \simeq \exp(-2)$. This tendency remains the same also for $\Lambda \neq 0$. As will be discussed in the next section, the physical wavelength corresponding to the peak mode can be modified by the succeeding amplification mechanism during the reheating.

\subsection{Upper bond on the gauge field production during inflation}\label{SSec:Upper bound in inflation}
In this subsection, using the formulae derived in Sec.~\ref{SSec:BR}, we compute a possible maximum amplitude of $\cA_h$, fulfilling the conditions (\ref{Exp:BRrho}) and (\ref{Exp:BReom}). During inflation, the Hubble parameter is directly related to the amplitude of the primordial tensor perturbation. Here, as a reference, we introduce expression of the primordial curvature perturbation and the tensor to scalar ratio for the slow-roll inflation (when the inflaton has the canonical kinetic term), given by
\begin{equation}
{\cal{P}}_{\zeta, {\rm sr}} \equiv \frac{ H^2}{8\pi^2\varepsilon_{\rm sr} \Mp^2}\,, \qquad r_{\rm sr} \equiv 16 \varepsilon_{\rm sr}\,, \label{Exp:PP}
\end{equation}
with
\begin{align}
    \varepsilon_{\rm sr} \equiv \frac{1}{18} \frac{V_{, \phi}^2}{\Mp^2 H^4}\,. \label{Def:eprilonsr}
\end{align}
Using these expressions, we rewrite the Hubble parameter as
\begin{align}
  &  \frac{H}{\Mp}
  = \frac{\pi}{\sqrt{2}} \sqrt{r_{\rm sr} {\cal{P}}_{\zeta, {\rm sr}}}\,. \label{Exp:HMp}
\end{align}
Only when the slow-roll approximation holds, the power spectrum of the primordial perturbation for the wavenumber which has crossed the Hubble scale  at $a=a_{\rm gen}$ is indeed described by Eqs.~(\ref{Exp:PP}). {Otherwise, ${\cal{P}}_{\zeta, {\rm sr}}$ and $r_{\rm sr}$ should be just understood as parameters which characterize the dynamics of inflation.} Inserting Eqs.~(\ref{Def:eprilonsr}) and (\ref{Exp:HMp}) into Eq.~(\ref{Cond:B}) and setting $F_{\rm gen}$ to 1, we obtain
\begin{align}
    B_h (\eta_{{\rm gen}}, k_{m, {\rm gen}}) \ll \frac{\Mp^2}{\sqrt{\Delta (\ln k)_{{\rm gen}}}} \sqrt{r_{{\rm sr}, {\rm gen}} {\cal{P}}_{\zeta, {\rm sr}, {\rm gen}}} \times {\rm min} \left[ 1,\,\, \frac{1}{D},\,\,\frac{1}{\sqrt{\alpha D}} \Bigl(\frac{f}{\Mp} \Bigr)^{1/2} r_{{\rm sr}, {\rm gen}}^{1/4} \right] \,. \label{Cond:Binf}
\end{align}
When the axion is the inflaton, a saturation of either the condition (\ref{Exp:BRrho}) or (\ref{Exp:BReom}) would imply a termination of inflationary expansion. Therefore, in going from Eqs~\eqref{Cond:B} to \eqref{Cond:Binf} we have imposed a stricter inequality.

In the monodromy case, numerical calculations give $D \simeq {\cal O}(1)$. For a larger value of $\alpha$, the condition (\ref{Exp:BReom}) tends to give a tighter constraint than the condition (\ref{Exp:BRrho}). {We emphasize again that, to generate gauge field with energy density close to the upper bound in Eq.~\eqref{Cond:Binf}, a large coupling between the inflaton and the gauge field is required, the amplitude of which is bounded by the slow-roll condition.}
As we have discussed in the previous subsection, only the modes $k/(aH) < |\xi|$ undergo tachyonic instability. As one can see in Eq.~(\ref{Def:B}), once these modes approach to the asymptotic value, the amplitude of the (hyper) magnetic field for each $k$ decays as $ 1/a^2 $. Therefore, at each moment, the peak wavenumber $k_m$ of $B_h$ is roughly given by the one which has just crossed $k/(aH) \simeq |\xi|$, i.e.,
\begin{align}
    \frac{k_{m, {\rm gen}}}{a_{{\rm gen}} H_{{\rm gen}}} \simeq \sqrt{2 \varepsilon_{\rm gen}} \alpha \frac{\Mp}{f}\,,   \label{Exp:kminf}
\end{align}
where we used Eq.~(\ref{Exp:xi}). Notice that $\varepsilon$ can deviate from $\varepsilon_{\rm sr}$, when the slow-roll approximation is violated. 

When we consider the gauge field generated around the moment when the fluctuations at the CMB scales have crossed the Hubble scale, the Hubble parameter and the slow-roll parameters should be compatible with the constraints from the CMB observation. For example, the tensor to scalar ratio and the amplitude of the primordial perturbation read 
$r <0.1 $ on $k={0.002}{\rm Mpc}^{-1}$ and ${\cal{P}}_\zeta \approx 2.2\times10^{-9}$
on $k=0.05{\rm Mpc}^{-1}$~\cite{Planck18}. Meanwhile, since the gauge field can generate the curvature perturbation which is highly non-Gaussian. As studied in Refs.~\cite{Barnaby:2010vf} and \cite{Barnaby:2011vw}, the perturbation generated by the gauge field should be subdominant relative
to the primordial curvature perturbations on the CMB scales.
This requirement provides $ |\xi_{\rm CMB}| < 2.37 $~\cite{Pajer:2013fsa}. 

On the other hand, these CMB constraints do not directly restrict the time evolution of the inflaton around the end of inflation (apart from the indirect constraint through $N_k$). Therefore, the parameters of the inflaton, which are repharased by  ${\cal P}_{\zeta, {\rm sr}}$ and $r_{\rm sr}$, and $\alpha$ in Eq.~\eqref{Cond:B}, do not necessarily refer to the CMB constraints.
For example, when we increase the value of $r_{\rm sr} \propto \varepsilon_{\rm sr}$, we can enhance the upper bound given in Eq.~(\ref{Cond:B}). In addition, the inverse dependence on $ \sqrt{\Delta\ln k} $ implies that, if the spectrum is peaked, the upper bound can be relaxed further.

\section{Gauge field production during reheating: Axion $=$ Inflaton}\label{Sec:production during axion reheating}

In the previous section, we have considered the generation of the gauge field during inflation, requesting that the slow-roll parameter $\varepsilon$ should be less than 1. In this section, we consider a generation of the gauge field during reheating, which takes place after inflation (for reviews, see, e.g., Refs.~\cite{Kofman:1997yn, Amin:2014eta, Lozanov:2019jxc}). During reheating, unlike during inflation, the velocity of the axion is not bounded by the slow-roll condition. In particular, the coherent oscillation of the axion can lead to a more efficient generation of the gauge field through the parametric resonance. In this section, we will show that such coherent oscillation of the axion (inflaton) leads to the resonant production of the gauge field, which can lead to a saturation of the upper bound on $B$ given in Eq.~(\ref{Cond:B}). 

\subsection{Resonant gauge field production}\label{SSec:resonant production}
At the end of inflation, the inflaton energy should be transferred to the one for the standard model particles. The coherent oscillation of the inflaton can realize this through parametric resonance. Similarly, the coherent oscillation of the axion can lead to the resonant production of the gauge field in our setup. A successful energy transfer from the coherently oscillating axion can be realized by enhancing either the rate of production or extending the duration of the parametric resonance phase. The former requires a large coupling between the axion and the gauge field (see e.g., Ref.~\cite{Adshead:2016iae}). Here, we consider the latter, which can lead to the efficient amplification of gauge field, even for $\alpha = {\cal O}(1)$.  

A sustainable resonance instability, in fact, can take place, when the time scale of the cosmic expansion is much longer than that of the oscillation, even just after the onset of the oscillation. When the axion is a subdominant component, this is possible if it is initially located in a potential region which is shallower than the quadratic potential, before the onset of the oscillation, as was argued, e.g., in Refs.~\cite{Soda:2017dsu, Kitajima:2018zco}. When the axion is the dominant component of the Universe, the condition for the delayed onset of oscillation is different. To illustrate this, here we consider the scalar potential with a dimensionless prescription. Using the dimensionless time coordinate $\tilde{t} \equiv mt$ and $\tilde{\phi} \equiv \phi/f$, we can express the background equation of motion as
\begin{align}
&\frac{\text{d}^2\tilde{\phi}}{\text{d} \tilde{t}^2} +  3\frac{H}{m}\frac{\text{d}\tilde{\phi}}{\text{d} \tilde{t}}+ \tilde{V}_{, \tilde{\phi}} = 0\,, \label{Eq:KGdl} \\
&  \left( \frac{H}{m} \right)^2  = \frac{1}{6} \left( \frac{f}{\Mp} \right)^2 \left[\left(\frac{\text{d}\tilde{\phi}}{\text{d} \tilde{t}} \right)^2  + 2 \tilde{V} \right]\,,  \label{Eq:Fdl}
\end{align}
and the equation of motion for $\cA_h$ with $h = \pm$ as
\begin{equation}
\frac{\text{d}\mathcal{A}_h}{\text{d} \tilde{t}^2} + \frac{H}{m}\frac{\text{d}\mathcal{A}_h}{\text{d} \tilde{t}} + \omega_h^2 \mathcal{A}_h = 0\,. \label{Eq:Adl}
\end{equation}
Here we have introduced the dimensionless (squared) frequency $\omega_h^2$ defined by
\begin{align}
   \omega_h^2 \equiv  \frac{k^2}{a^2m^2} -  h \alpha \frac{k}{ma}\left(\frac{\text{d}\tilde{\phi}}{\text{d} \tilde{t}}\right) \,.   
\end{align}
The strength of the Chern-Simons coupling is sometimes expressed by the combination $\alpha \Mp/f$, in particular when discussing the end of the large field inflation, where $\phi$ is of ${\cal O} (\Mp)$. Here and hereafter, we only consider the case where the amplitude of $\phi$ remains ${\cal O}(f)$ and the coupling (which appears in the Lagrangian) amounts to $\alpha (\phi/f) \simeq \alpha$. Therefore, we use $\alpha$ to characterize the strength of the coupling. 

\begin{figure}
    \centering
    \includegraphics[width=1.\textwidth]{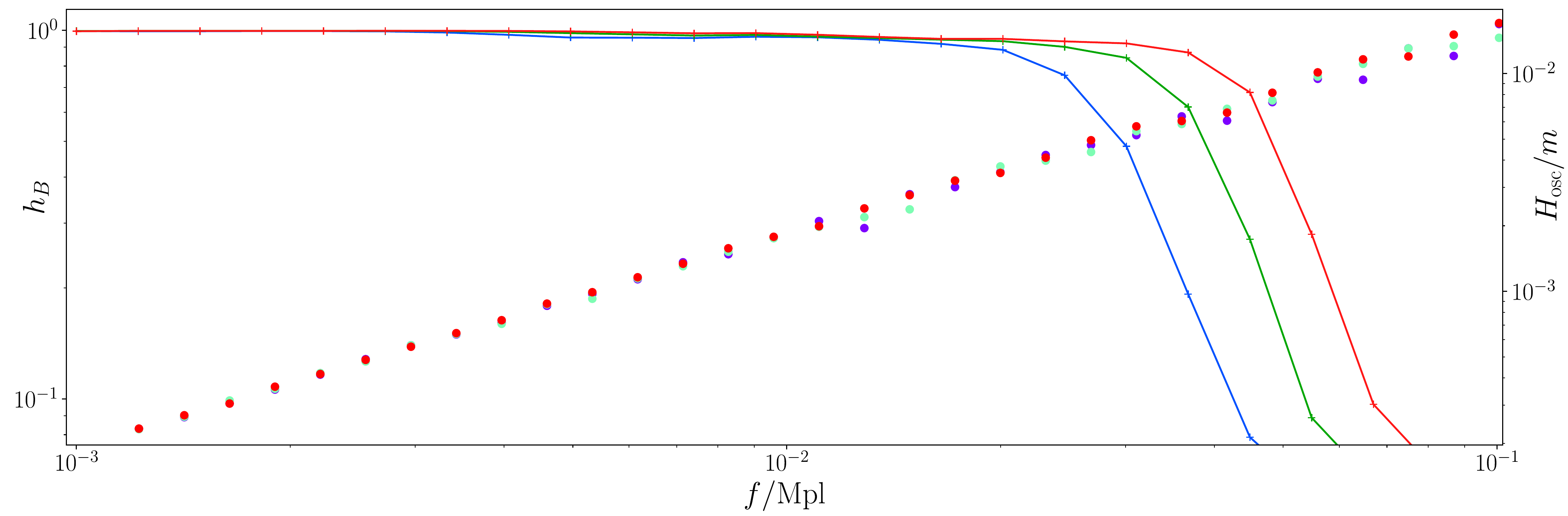}    \caption{This plot shows the value of $H_{\rm osc}/m$ (circular markers) and the normalized helicity fraction $h_B$ (solid lines) defined in $\eqref{Def:Helicity Fraction}$ as a function of $f/{\Mp}$. Here, $H_{\rm osc}$ denotes the Hubble parameter, when $\varepsilon$ has reached peak value after the slow-roll evolution. The violet, green and red locus of markers correspond to the potential parameter $p=2,4.9$ and $8$. The blue, green, and red lines represent $h_B$ computed at the end of 50 oscillation cycles with the potential parameter $p=2$ and $\alpha = 0.8,\, 0.9\, {\rm and}\, 1$, respectively. 
    }\label{Fg:Hoscf}
\end{figure}
When the axion starts oscillating coherently after rolling down the plateau region, the two terms in the square brackets of Eq.~(\ref{Eq:Fdl}) both become ${\cal O}(1)$. In this paper, we denote the time at which the axion velocity $|\D \tilde{\phi}/\D \tilde{t} |$ reaches its peak value as the onset of oscillation, which corresponds to the beginning of the reheating epoch in the setup of this section. Therefore, we can roughly estimate the Hubble parameter around the onset of the oscillation as (see also Ref.~\cite{Lozanov:2017hjm})
\begin{equation}
    \frac{H_{\rm osc}}{m} \simeq \frac{1}{\sqrt{6}}\frac{f}{\Mp}\,.   \label{Exp:Hoscinf}
\end{equation}
This indicates that the onset of oscillation significantly delays for $f/\Mp \ll 1$\footnote{Equation (\ref{Exp:Hoscinf}) applies, only when two terms in the square brackets of Eq.~(\ref{Eq:Fdl}) are of ${\cal O}(1)$ at the onset of the oscillation. For natural inflation~\cite{Freese:1990rb}, where $\tilde{V} \equiv V/(mf)^2$ is given by $\tilde{V}= 1 - \cos \tilde{\phi}$, the slow-roll condition requires $f > \Mp$ and $\phi_{\rm osc} = {\cal O}(\Mp)$. Therefore, Eq.~(\ref{Exp:Hoscinf}) cannot be applied.}. By numerically solving Eqs.~(\ref{Eq:KGdl}) and (\ref{Eq:Fdl}), this can be confirmed as shown in Fig.~\ref{Fg:Hoscf}, where we considered the potential given by
\begin{equation}
    V(\phi) = \frac{(mf)^2}{2} \left[  1 - \frac{1}{(1+\tilde{\phi}^2/p)^p} \right] \equiv (mf)^2 \tilde{V}(\tilde{\phi}) \,.\label{Pure Natural Potential}
\end{equation}
In Fig.~\ref{Fg:Hoscf}, we present the values of $H_{\rm osc}/m$ for different $f/M_{\rm pl}$ and potential parameters $p=2,\,4.9,$ and $\,8$, with the initial condition $\tilde{\phi}_{\rm i}=6$.

\begin{figure}
    \centering
    \includegraphics[width = 1.\textwidth]{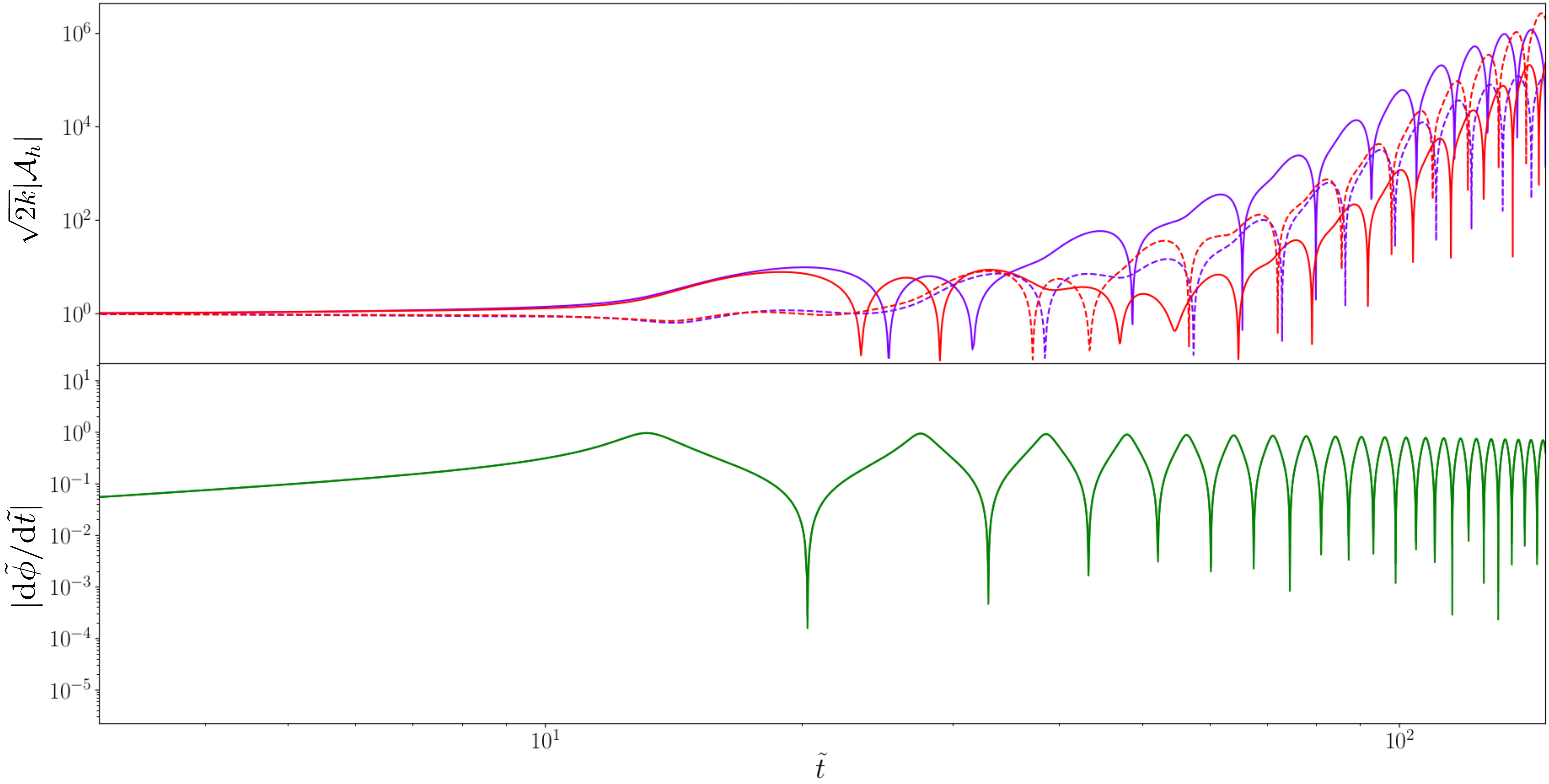}
    \caption{These plots show the time evolution of the gauge field $\cA_h$ (top panel) and the time derivative of the axion inflaton field (bottom panel).
Here, the values of parameters are $f/\Mp = 0.01$, $\alpha=1$, and $p=2$. The evolution of ${\cal A}_h$ with wavenumbers $k/(ma_{\rm osc})=0.25$ (red) and $0.2$ violet are displayed in the upper panel, where the solid and dashed curves correspond to the $h=-$ and $h=+$ polarization modes, respectively.
    }
\label{Fg:evreh}
\end{figure}

\begin{figure}
    \centering
    \includegraphics[width = 1.\textwidth]{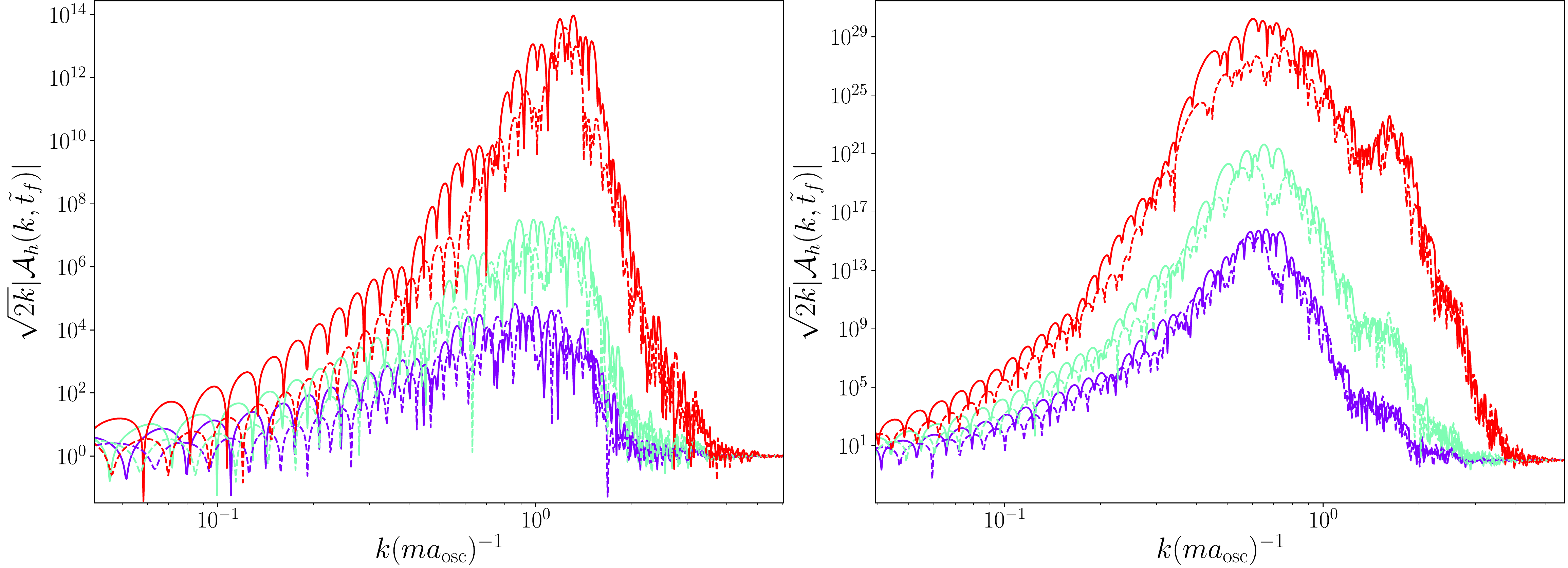}
    \caption{The spectra of the gauge field after 50 oscillation cycles of the background axion during reheating after pure natural inflation are displayed in the left ($f/\Mp=0.05$) and right ($f/\Mp=0.01$) panels. The solid and dashed lines show $h=-$ and $h=+$, respectively. The violet, green and red lines correspond to the dimensionless coupling constant $\alpha = 1.5, 2$ and $3$, respectively.
    }\label{Fg:spreh}
\end{figure}

For $H_{\rm osc}/m \ll 1$, the cosmic expansion does not disturb the exponential growth of the gauge field \cite{Soda:2017dsu, Kitajima:2018zco}. In this case, since $\omega_h^2$ becomes a quasi periodic function, Eq.~(\ref{Eq:Adl}) can be well approximated by the Hill's equation. In the Hill's equation, the position of the resonance band is characterized by the amplitude of the non-oscillatory contribution, the first term of $\omega_h^2$, and the growth rate is characterized by the amplitude of the oscillatory contribution, the second term of $\omega_h^2$ (see e.g., Ref.~\cite{Fukunaga}). In particular, for the Mathieu equation, whose periodic function (in $\omega_h^2$) is given by the sinusoidal function, the wavenumber for the first resonance band reads $k/(am) \sim 1/2$.

Figure \ref{Fg:evreh} shows the time evolution of $\cA_h$~(top panel) and ${\rm d} \tilde{\phi}/ ({\rm d} \tilde{t})$~(bottom panel) for $f/\Mp = 0.01$, $\alpha=1$, and $p=2$.
Towards the end of inflation (around $\tilde{t} \simeq 10$), the velocity of the axion increases, leading to the tachyonic instability of the gauge field. Therefore, an amplitude difference arises between the two polarizations of the gauge field~(the amplitude difference between solid and dotted lines), resulting in non-zero helicity density.  

Once the axion commences to oscillate after $\tilde{t} \simeq 10$ in Fig.~\ref{Fg:evreh}, the gauge field grows exponentially through the parametric resonance instability. Since $H_{\rm osc}/m$ is much smaller than 1, the amplitude of ${\rm d} \tilde{\phi}/ ({\rm d} \tilde{t})$ does not significantly decrease in the time scale of the background oscillation. As a result, the gauge field keeps on growing without a significant disturbance due to the cosmic expansion.
The parametric resonance amplifies the two different polarization modes indistinguishably. The helicity fraction $h_B$, defined as
\begin{equation}
    \label{Def:Helicity Fraction}
    h_B \equiv \frac{\int {\text{d}^3\mathbf{k}}k(|\mathcal{A}_+|^2-|\mathcal{A}_-|^2)}{\int {\text{d}^3\mathbf{k}}k(|\mathcal{A}_+|^2+|\mathcal{A}_-|^2)}\,,
\end{equation}
grows due to the tachyonic instability and approaches a non-vanishing constant value soon after the onset of the oscillation. The helicity fraction $h_B$ is related to $\langle \mathfrak{H} \rangle $ roughly as $\langle \mathfrak{H} \rangle \sim h_B \Delta \ln k_m k_m^4 |{\cal{A}}_h|^2$, where $|{\cal{A}}_h|$ is the amplitude of the dominant polarization mode evaluated at the peak wavenumber $k_m$. Therefore, the amplification of $|{\cal{A}}_h|$ due to the subsequent parametric resonance keeps on enhancing the helicity density $\langle \mathfrak{H} \rangle $, while preserving $h_B$ (unless the non-linear interaction washes out the helicity after the saturation ~\cite{Adshead:2015pva}). Figure~\ref{Fg:Hoscf} shows the value of the helicity fraction at the end of $50$ oscillation cycles of the inflaton as a function of $f/\Mp$ and the coupling constant $\alpha$. For $H_{\rm osc}/ m \ll 1$, the tachyonic instability persists longer, before the parametric resonance sets in. This leads to the fully circularly polarized (hyper) magnetic field.

Figure \ref{Fg:spreh} shows the spectrum of $\cA_h$ for $f/\Mp = 0.05$ (left) and $ 0.001$ (right) with $\alpha = 1.5$ (violet), $\alpha= 2$ (green) and $\alpha = 3$ (red), after 50 oscillation cycles of the background axion.
Here, the value of $p$ is set to $p=2$ and the initial condition $\tilde{\phi}_{\rm i} = 3$ is adopted. As we have discussed, a smaller $f$ leads to a more prominent growth due to the resonance instability without being disturbed by the cosmic expansion.
The spectrum has a peak around $k/(am) \sim 1/2$ because of the first resonance band.
Since tachyonic amplification is less efficient for small values of $\alpha$, the generated $h_B$ is smaller, as can be seen from Fig.~\ref{Fg:Hoscf}.

\begin{figure}
    \centering
    \includegraphics[width = 1.\textwidth]{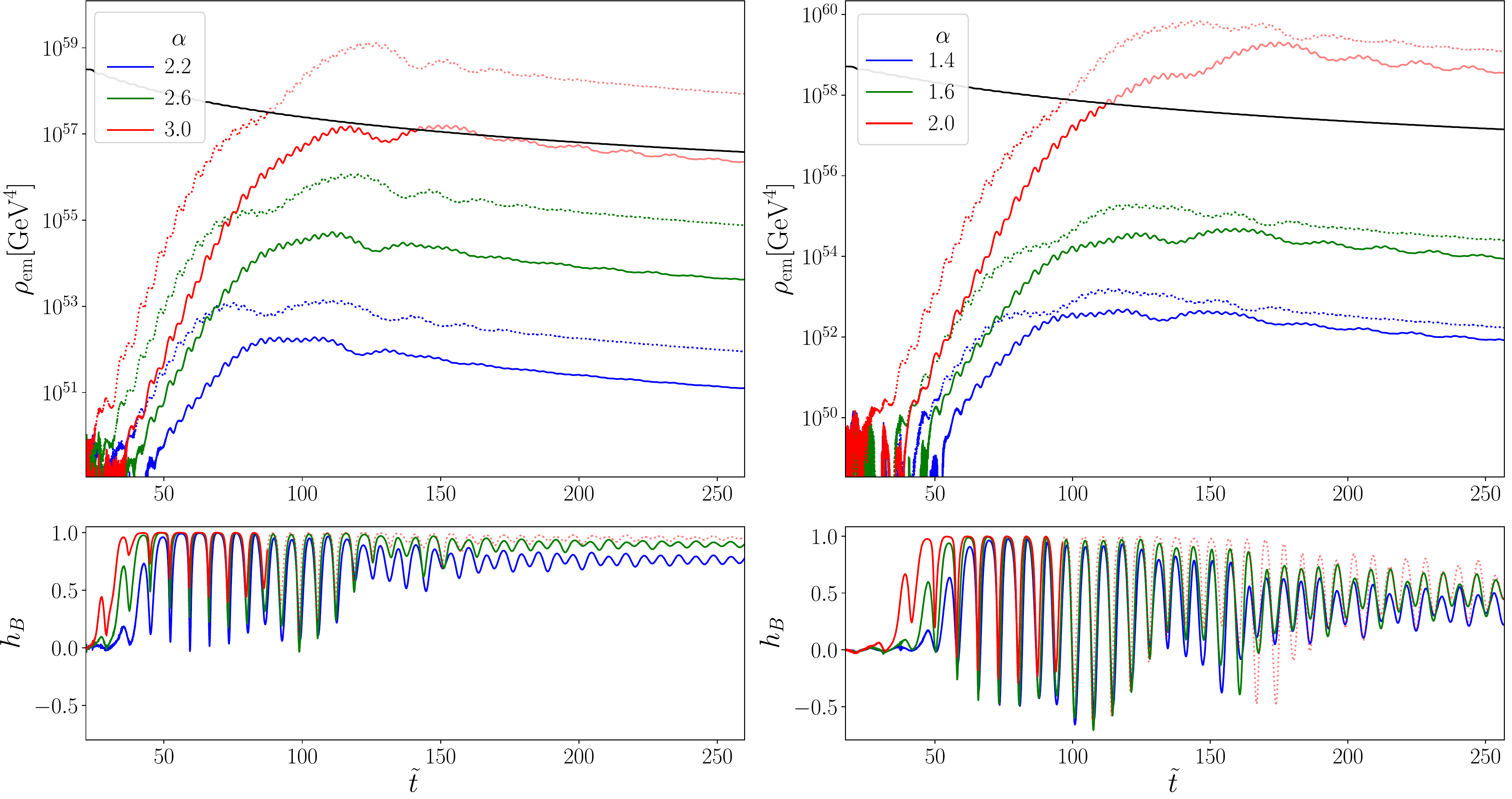}
    \caption{
    The top panels show the time evolution of the energy density of the U(1) gauge field. The values of the coupling constant corresponding to the different colors of the lines are shown in the legend. The solid black lines shows the evolution of the energy density of the inflaton $\rho_\phi$. Here, we chose the decay constant as $f/\Mp = 0.1$ (left) and $f/\Mp=0.05$ (right), respectively, setting the initial $\rho_{\phi}$ as {$\rho_\phi = (5.1\times 10^{14}{\rm GeV})^4$}. Then, the corresponding axion masses differ as $m = 6.1\times 10^{11}{\rm GeV}$ (left) and $1.2\times 10^{12}{\rm GeV}$ (right). In the bottom panels, the evolution of the helicity fraction $h_B$ is shown.}
    \label{Fg:rhoreh}
\end{figure}

Figure~\ref{Fg:rhoreh} shows the time evolution of the energy density for the gauge field, $\rho_{\rm em}$, given by the sum of Eqs.~(\ref{Exp:RhoE}) and (\ref{Exp:RhoB}). To compute the energy density of the generated gauge field, we adopted the adiabatic regularization. A larger value of $\alpha$ and a smaller value of $f$ lead to a more prominent growth of the gauge field. As is shown, $\rho_{\rm em}$ can become comparable with $\rho_{\phi}$ (the black curves) {for some sets of the parameters}. When we properly follow the non-linear dynamics, the resonant growth of the gauge field would terminate when $\rho_{\rm em} \simeq \rho_{\phi}$. The spurious successive growth in Fig.~\ref{Fg:rhoreh} is caused, since we ignored the backreaction of the gauge field production. 

The bottom panel in Fig.~\ref{Fg:rhoreh} outlines the time evolution of the helicity fraction $h_B$. We can see that $h_B$ rapidly grows in the early stage, where the magnetic field evolves through the tachyonic instability.
When the parametric resonance halts, the oscillation of $h_B$ also attenuates and, finally, a non-zero $h_B$ is left in the produced gauge fields.
 
\subsection{Order estimation of the gauge field production during reheating}\label{SSec:Order reheating}
For a general Hill's equation, the structure of the resonance band can be different, depending on the periodic function in $\omega_h^2$ (see, e.g., Ref.~\cite{Fukunaga}). However, unless the oscillation period significantly deviates from $ {\cal O}(1)/m$, we obtain the peak wavenumber as
\begin{equation}
    \frac{k_{m,\, {\rm gen}}}{m a_{{\rm gen}}} = {\cal O} (1)\,, \label{Exp:kmreh}
\end{equation}
and the width of the resonance band as $\Delta \ln k = {\cal O}(\alpha)$. Since the characteristic time scale of $\cA_h$ is roughly of the order of the oscillation time scale $m$, we obtain $D = {\cal O}(1)$ using Eq.~(\ref{Exp:kmreh}). Inserting these expressions into Eq.~(\ref{Cond:B}) and setting $F_{\rm gen}$ to $1$, we obtain the upper bound on the (hyper) magnetic field $B_h$ as
\begin{align}
    B_h (t_{\rm gen},\, k_{m,\, {\rm gen}}) \alt \frac{H_{{\rm gen}} \Mp}{\sqrt{\alpha}} \times {\rm min} \left[1,\,\, \frac{1}{\sqrt{\alpha}}\right] \,. \label{Cond:Breh}
\end{align}
Here, assuming that $H_{\rm osc} \simeq H_{{\rm gen}}$, which can be verified for $f/\Mp \ll 1$, we used $m/H_{\rm gen} \simeq m/H_{\rm osc} \simeq \Mp/f$. The upper bound which is proportional to $1/\sqrt{\alpha}$ comes from Eq.~(\ref{Exp:BRrho}) and the other one, which is proportional to $1/\alpha$, comes form Eq.~(\ref{Exp:BReom}). Notice that since both $d {\cal A}_h/d \tilde{t}$ and $(k/am) {\cal A}_h$ amount to ${\cal O}(1) \times {\cal A}_h$ for the gauge field generated through the parametric resonance, $\rho_{\rm E}$ and $\rho_{\rm B}$ are almost comparable. Among the cases we address in this paper, this is common as long as the parametric resonance is the dominant mechanism for the gauge field production. 

Equation (\ref{Cond:Breh}) states that the gauge field production with a small $\alpha$ can enhance the upper bound of $B_h$ at the generation as a consequence of having the narrow spectrum and relaxing the condition (\ref{Exp:BReom}). Nevertheless, the overall factor $1/\sqrt{\alpha}$ is cancelled in the helicity density and it does not lead to a net enhancement of the helical magnetic field at present (see Sec.~\ref{SSSec:inverse cascade}). 

We derive Eq.~(\ref{Cond:Breh}), assuming that the backreaction of the gauge field production does not significantly modify the evolution of the background spacetime and the axion. During reheating, unlike during other epochs whose dynamics are more strictly constrained, a sizable backreaction does not immediately create a problem. However, a more expensive computation scheme, such as lattice simulation, is required to properly follow the non-linear dynamics. Furthermore, when the inhomogeneous mode of the axion becomes non-negligible as $|\delta \phi| \simeq |\phi|$, the two polarization modes can interact with each other, leading to a reduction of the generated helicity density. In this paper, we do not consider such a parameter range.

Assuming that the slow-roll condition is satisfied up until the onset of the oscillation, we can rewrite Eq.~(\ref{Cond:Breh}) as
\begin{align}
    B_h (t_{\rm gen},\, k_{m,\, {\rm gen}}) \alt \Mp^2 \left( r_{{\rm CMB}} {\cal P}_{\zeta, {\rm CMB}} \over \alpha \right)^{1/2}  \times {\rm min} \left[1,\,\, \frac{1}{\sqrt{\alpha}}\right] \,, \label{Cond:Breh2}
\end{align}
where ${\cal P}_{\zeta, {\rm CMB}}$ and $r_{{\rm CMB}}$ denote the dimensionless spectrum of the primordial curvature perturbation and the tensor to scalar ratio at the CMB scale. Here, we used $H_{\rm osc} \simeq H_{\rm CMB} \simeq \Mp \sqrt{r_{{\rm CMB}} {\cal P}_{\zeta, {\rm CMB}} }$, where $H_{\rm CMB}$ denotes the Hubble parameter at the time when CMB scale fluctuation had crossed the Hubble scale.

Compared to the setup in Sec.\ref{Sec:gauge field generation during inflation}, the gauge field production during reheating can be rather efficient, especially for $f/\Mp \ll 1$. As was shown in the previous subsection, the gauge field can grow exponentially through the parametric resonance, so that the energy density of the gauge field becomes comparable with that of the axion, even when the dimensionless coupling normalized by the typical value of the present setup, $\alpha$, is ${\cal O}(1)$. Therefore, in this case, saturating the upper bound in Eq.~(\ref{Cond:Breh}) is possible.
However, the present coherent length of the magnetic field generated during reheating is much smaller than cosmological scales, when it evolves adiabatically.
Nevertheless, since the generated gauge field has non-zero helicity density,  the subsequent inverse cascade may allow us to overcome this problem. We will discuss the evolution after the generation in Sec.~\ref{SSSec:inverse cascade}.

Here, we comment on the effect of the particle creation during reheating.
When the onset of the oscillation delays, taking $H_{\rm osc}/m \ll 1$, production of other particles, including SM particles, will also become efficient. If the Universe comes to be dominated by the plasma of such particles created before the generation of the gauge field, a large conductivity will disturb the gauge field production (a relevant discussion will be given in Sec.~\ref{Sec:post inflationary gen}). Here, we have simply assumed that the coupling between the axion and the gauge field is large enough such that the energy transfer from the axion to the gauge field is completed, prior to the creation of plasma. 

\section{Gauge field production during inflation: Axion $\neq$ Inflaton}\label{Sec:spectator production}
In the previous two sections, we have discussed the case where the axion drives the inflationary expansion. During inflation, we can generate the gauge field whose coherent length largely exceeds the Hubble scale, while the possible coupling between the inflaton and the gauge field is tightly constrained by the slow-roll condition. During reheating, we can more efficiently enhance the gauge field through the parametric resonance, while the coherent length is at most the Hubble scale at that time. In this section, we investigate whether we can resonantly enhance the gauge field during inflation so that the inflationary expansion stretches the physical wavelength by considering an axion which is not the inflaton, but a spectator field.  

\subsection{Resonant gauge field production from spectator axion} \label{SSec:unharmonic setup}
When the axion is not the inflaton, the impact of the axion evolution on the inflationary expansion is negligible. Therefore, it is enough to solve Eqs.~(\ref{Eq:KGdl}) and (\ref{Eq:Adl}) for a fixed inflationary spacetime. As was discussed in Ref.~\cite{Kitajima:2018zco}, when the axion is not the dominant ingredient of the Universe and the axion was initially located at a shallow potential region, the onset of the oscillation can be roughly estimated by  
\begin{equation}\label{Exp:Hosc_PN}
    H_{\rm osc} \sim \sqrt{\left| \frac{V_{, \phi}(\phi_{\rm i})}{\phi_{\rm i}} \right|} \,,
\end{equation}
where $\phi_{\rm i}$ denotes the initial value of the axion. In particular, when the axion is initially located at a region where the gradient of the potential is much shallower than the one for $\phi^2$, $H_{\rm osc}$ becomes much smaller than $m$, where $m$ is defined by the curvature of the potential minimum.

\begin{figure}
	\centering
	\includegraphics[width=1.\linewidth]{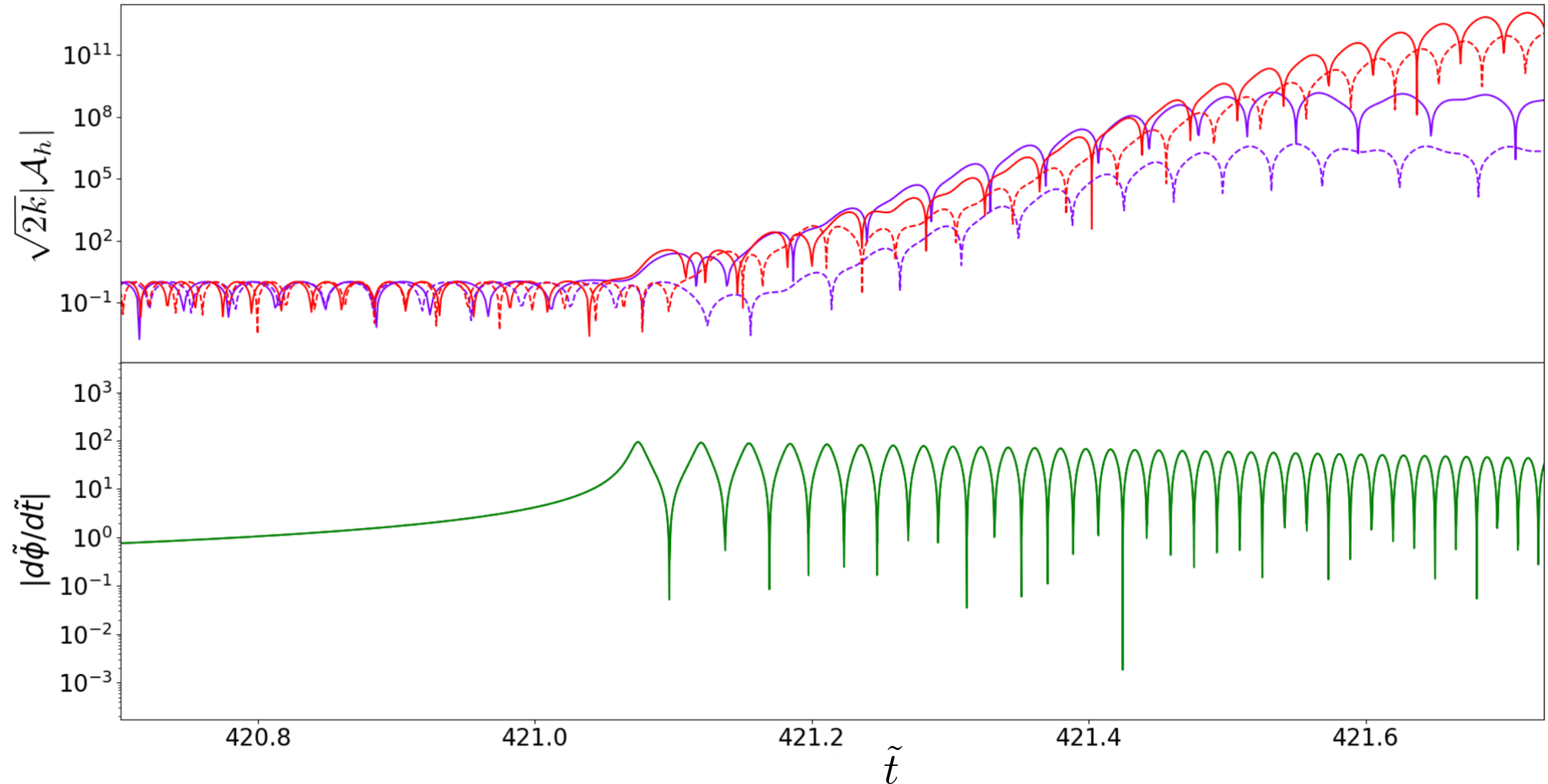}
	\caption {The evolution of the gauge field $\cA_h$ (top panel) and the spectator axion (bottom panel). The solid and dashed lines in the upper panel correspond to the $h=\pm$ helical modes, respectively. The violet and red lines correpond to wavenumbers $k/(ma_f)=0.17$ and $ 0.27$, respectively. In the bottom panel, the evolution of $|{\D \tilde{\phi}}/{\D \tilde{t}}|$ is shown as solid green curve. Here we set the potential exponent $p=6$ and the coupling constant $\alpha = 5$.}
	\label{fig:inf_spectator_mode}
\end{figure}

As an example, here we study the time evolution of the gauge field which is coupled with the spectator axion whose potential is given by Eq.~(\ref{Pure Natural Potential}). The Hubble parameter during inflation would evolve as $H/m = (a/a_{H=m})^{-\varepsilon}$, where $a_{H=m}$ denotes the scale factor when $H \simeq m$. Here, assuming that the contribution of the axion to the background energy density is negligible, we simply take the slow-roll parameter $\varepsilon\,(\ll 1)$ as a given free parameter.

Figure~\ref{fig:inf_spectator_mode} shows the time evolution of the background axion and the gauge field $\cA_h$ with $h=\pm$. In this figure, we set $p = 6$ and $\tilde{\phi}_{\rm i}=4$.
Here, oscillation occurs at $H_{\rm osc}/m =0.015$.
Similar to the situation discussed in the previous section, the gauge field first grows due to tachyonic instability and subsequently, through parametric resonance. When the onset of the oscillation significantly delays, the exponential growth due to the parametric resonance would persist up until the energy density of the gauge field becomes roughly comparable to that of the axion.

\begin{figure}
	\centering
	\includegraphics[width=1.\linewidth]{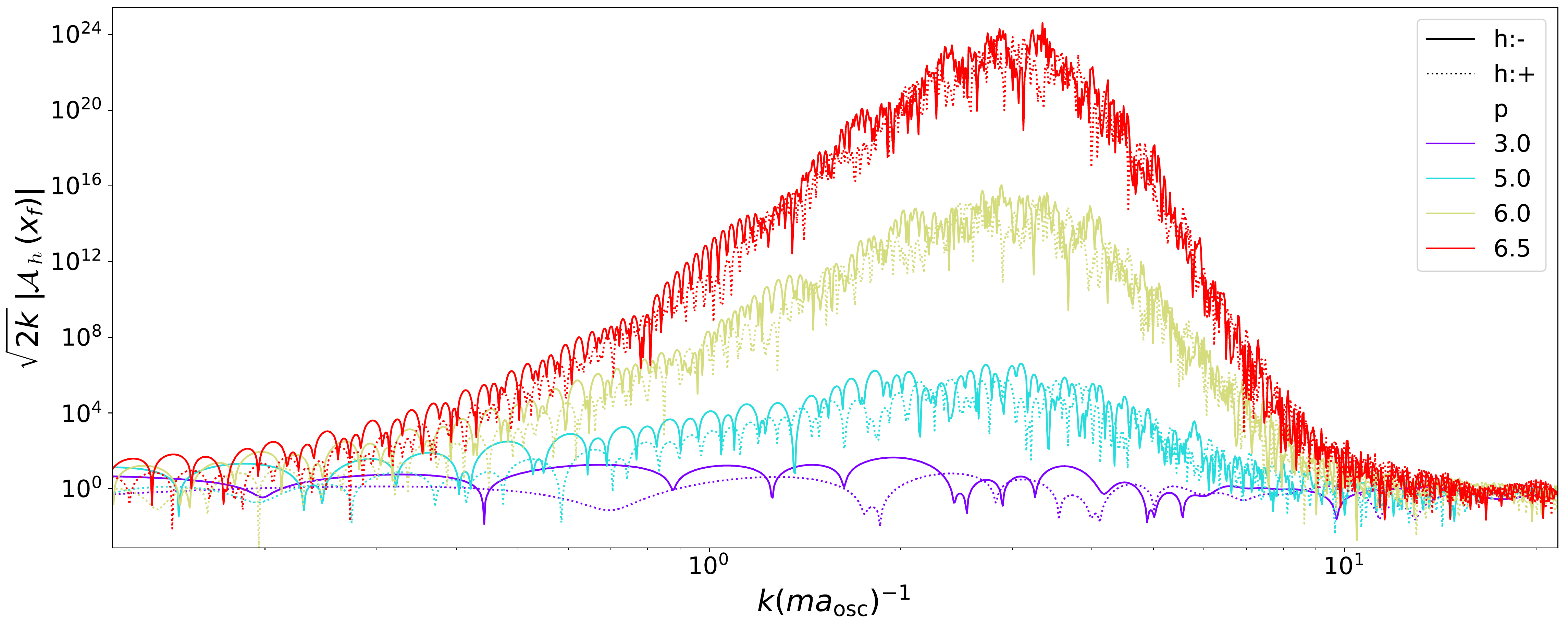}
	\caption{The curves are the spectra for exponent $p$ as labeled and $\alpha =5$. The horizontal axis corresponds to the wave number $k/(ma_{\rm osc})$, where $a_{\rm osc}$ denotes the scale factor at onset of oscillation.}
	\label{fig:spectra-alpha-5}
\end{figure}

\begin{figure}
	\centering
	\includegraphics[width=1\linewidth]{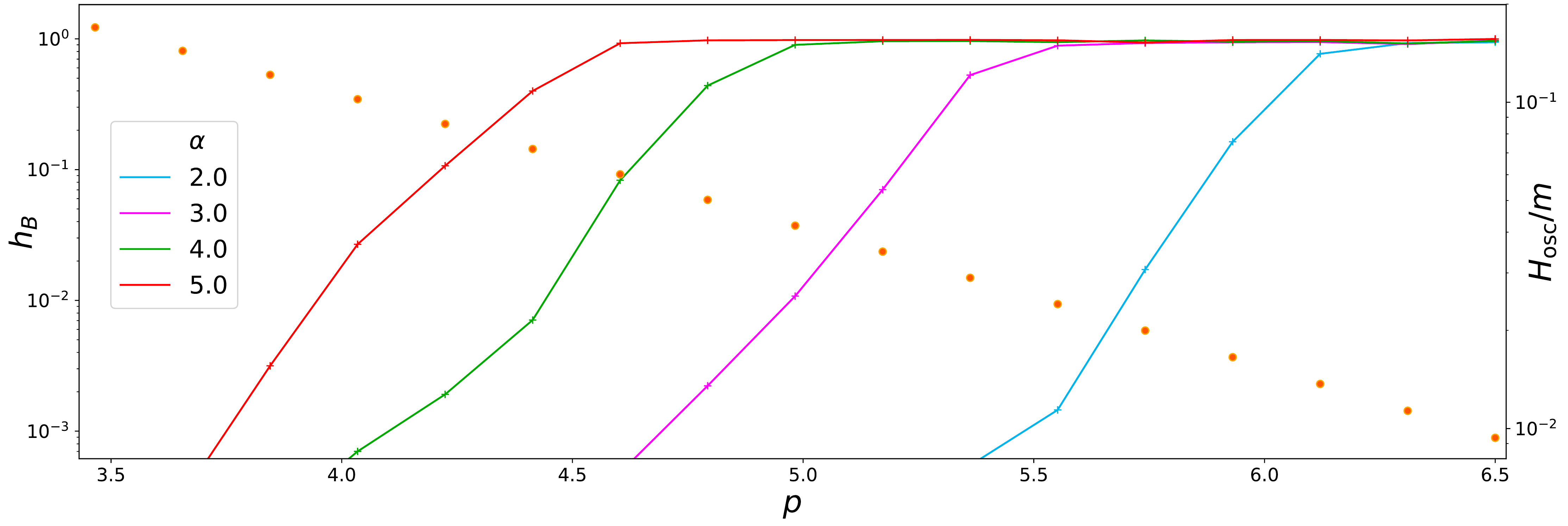}
	\caption{This plot shows the helicity fraction (solid lines) $h_B$ and $H_{\rm osc}/m$ (circular markers) for different values of the potential parameter $p$. The blue, magenta, green and red curves correspond to $ \alpha = 2,3,4$ and $5 $.}
	\label{fig:peak-hel-vs-p-alpha-1-2-3-4-5}
\end{figure}

For a fixed value of initial condition $\tilde{\phi}_{\rm i}, $ the potential exponent $p$ determines $ H_{\rm osc}/m $.
Figure~\ref{fig:spectra-alpha-5} represents the dependence of the generated gauge spectra on the potential exponent~$p$. As $p$ increases, since the potential (in the plateau region) becomes shallower, the onset of the oscillation delays more. As already discussed, the delay results in the efficient resonant amplification. The resultant amplitude becomes larger for a larger $p$.  In Fig.~\ref{fig:peak-hel-vs-p-alpha-1-2-3-4-5}, we plot the dependence of the helicity fraction $h_B$ (for different $\alpha$) and $H_{\rm osc}/m$ on $p$.
A smaller value of $H_{\rm osc}/m$ implies that the Hubble friction is less effective and that the tachyonic amplification leads to a larger value of $h_B$. 

\subsection{Order estimation of the gauge field production from spectator axion}
We have shown that, likewise during reheating, the spectator axion can resonantly enhance the gauge field during inflation just after the onset of the oscillation. In particular, when the axion was located at a shallow potential region before the onset of the oscillation, the gauge field can grow until $\rho_{\rm em}$ catches up with $\rho_{\phi}$. 

As was argued in the previous section, since we obtain $D= {\cal O}(1)$ and $\Delta \ln k = {\cal O}(\alpha)$ for the resonantly generated gauge field, Eq.~(\ref{Cond:B}) reads
\begin{align}
    B_h (\eta_{{\rm gen}}, k_{m, {\rm gen}}) \alt \frac{H_{{\rm gen}} \Mp}{\sqrt{\alpha}} \sqrt{F_{\rm gen}} \times {\rm min} \left[ 1,\,\,\,\,\frac{1}{\sqrt{\alpha}} \right] \,, \label{Cond:Bspc}
\end{align}
where we used $\sqrt{F_{\rm gen}} \simeq (mf)/(H_{\rm gen} \Mp)$. Similar to the gauge field production during reheating, saturating the upper bound here does not immediately pose an issue for the background inflationary dynamics. The maximum amplitude of the seed magnetic field is bounded by the amount of the possible energy source, $\sqrt{F_{\rm gen}}\, H_{\rm gen} \Mp \sim \sqrt{\rho_{\phi, {\rm gen}}}$. The peak wavenumber of the gauge field at the generation is again determined by the resonance band and is given by Eq.~(\ref{Exp:kmreh}). When the generation takes place during inflation, the physical wavelength is stretched afterwards by the inflationary expansion, while the (hyper) magnetic field is diluted unless there is another source to sustain it (for a more detailed discussion, see Sec.~\ref{Sec:present}). 

\section{Resonant magnetic field generation in a plasma filled Universe}
\label{Sec:post inflationary gen}
When an axion commences to oscillate after reheating, it may potentially generate the gauge field during radiation domination or matter domination. In this section, we investigate this possibility, considering the radiation dominant Universe.

\subsection{Electromagnetic field in plasma}
During reheating, the coherent oscillation of the inflaton leads to a creation of the plasma particles. In the presence of conducting plasma, the equation of motion for $\cA_h$ acquires an extra friction term as
\begin{equation}
\label{Post inflationary EM EOM}
 \frac{\D^2 \mathcal{A}_h}{\D \tilde{t}^2 } + \left(\frac{ H}{m}+\frac{4\pi\sigma}{m}\right) \frac{\D \mathcal{A}_h}{\D \tilde{t} } +  \omega_h^2 \mathcal{A}_h = 0 \,,
\end{equation}
where $\sigma$ denotes the conductivity of the Universe. For a consistent history of the Universe, the axion has to be subdominant, when it starts oscillating. The mass of the axion can acquire the thermal correction through the interaction with the Yang-Mills instanton (see, e.g., Refs.~\cite{Berkowitz:2015aua, Borsanyi:2016ksw}). Taking the thermal correction into account is left for future works.

The conductivity $\sigma$ depends on the temperature as
\begin{equation}
\label{Conductivity}
\sigma(T) = 
\begin{cases}
  10{\rm GeV}\left(\frac{T}{\rm GeV}\right)\,, & \text{for\,\,} T\gg m_e \,, \\
  3.2\times 10^4  {\rm GeV}\left(\frac{T}{\rm GeV}\right)^{\frac{3}{2}}\,,   &\text{for\,\,} T\ll m_e\,.
\end{cases}
\end{equation}
The first expression is valid when the temperature is greater than the electron mass $m_e$~\cite{Baym:1997gq}.
At lower temperatures, the electrons are no longer relativistic and the conductivity is given by Coulomb scattering of electrons in a fully ionized gas, as given by Spitzer-Harm resistivity~\cite{Spitzer and Harm}.

\subsection{Can the magnetic field be resonantly produced in plasma?}

According to Eq.~\eqref{Post inflationary EM EOM},
the conductivity can disturb the motion of oscillation similar to the Hubble friction.
In the preceding sections, we have shown that, during inflation and reheating~(in the limit of $\sigma=0$), the gauge field can be resonantly produced for $H_{\rm osc} \ll m$, since the Hubble friction is negligible in the time scale of the oscillation.
Similarly, for 
\begin{equation}
    \sigma(T) \ll m\,,  \label{Eq:sigma}
\end{equation}
the resonant amplification can take place\footnote{To be more precise, the exponential growth can take place, when the Floquet exponent $\mu$ satisfies ${\rm Re}[\mu] > \sigma/m $. Here, assuming that ${\rm Re}[\mu] \alt {\cal O}(1)$, which is typically the case for $\alpha \alt {\cal O}(1)$, we simply consider the condition (\ref{Eq:sigma}).}. (Another scenario to circumvent the large conductivity was discussed in Ref.~\cite{Fujita:2019pmi}.)

\begin{figure}
    \centering
    \includegraphics[width = 1.\textwidth]{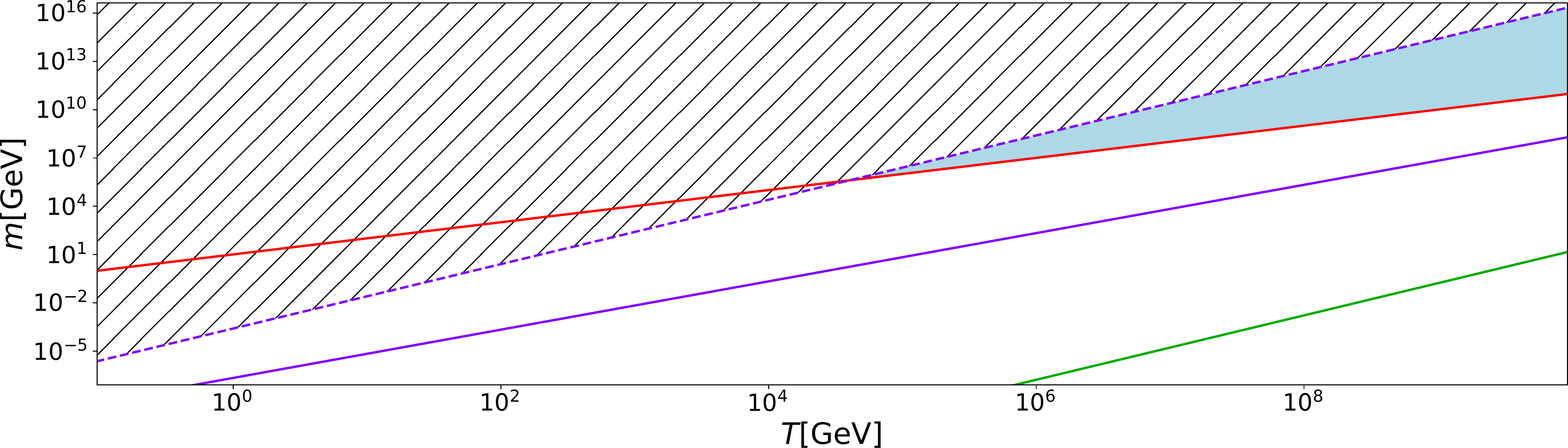}
    \caption{
    The region above the red curve satisfies the condition (\ref{Eq:sigma}) and the region below the violet solid curve satisfies the condition (\ref{constraint on axion energy density rad epoch}) for $f= 1 {\rm TeV}$. As is shown, there is no viable parameter range which satisfies both Eqs.~(\ref{Eq:sigma}) and (\ref{constraint on axion energy density rad epoch}). The region shaded in light blue below the violet dashed line satisfies the relaxed condition (\ref{constraint on mass range imposed by R.E.D.}) and Eq.~(\ref{Eq:sigma}).
    }
    \label{fig:mass range}
\end{figure}

Here, in order to avoid spoiling the cosmological expansion history, the energy density of the axion which sources the gauge field is bounded by above. The simplest possibility is that, after the generation of the gauge field, the energy density of the residual axion evolves adiabatically as $\rho_\phi \propto 1/a^3$. Then, since the relic abundance of the axion should not exceed the total dark matter abundance, we obtain
\begin{equation}
\label{constraint on axion energy density rad epoch}
\begin{aligned}
m \leq 4\times10^{-8}{\rm GeV}\left( \frac{10^3 {\rm GeV}}{f} \right) \left( g_{* s, {\rm osc}} \over g_{* s, 0} \right)^{\frac{1}{2}} \left( \frac{T_{\rm osc}}{\rm GeV} \right)^{\frac{3}{2}}\,.
\end{aligned}
\end{equation}

Figure \ref{fig:mass range} shows range of axion mass for each $T_{\rm osc}$ that satisfies both Eqs.~(\ref{Eq:sigma}) and (\ref{constraint on axion energy density rad epoch}). The former is shown by the red solid line and the latter is shown by the violet solid line. Here, we choose the decay constant $f$ as $f = 1 {\rm TeV}$ (the constraints on the coupling with the photon for heavy axions were discussed e.g., in Ref.~\cite{Fukuda:2015ana}). Figure \ref{fig:mass range} shows that there is no parameter range which satisfies both Eqs.~(\ref{Eq:sigma}) and (\ref{constraint on axion energy density rad epoch}).

One may come up with a more contrived setup where the axion decays into dark sector component after the generation of the magnetic field~(see e.g. Ref.~\cite{Agrawal:2018vin}). In this case, the condition on the energy density of the axion is that, at the onset of oscillation, it does not exceed the radiation energy density, which gives

\begin{align}
    \label{constraint on mass range imposed by R.E.D.}
 \rho_{\phi, {\rm osc}} \simeq  (mf)^2 <   3 \Mp^2 H_{\rm osc}^2\,.
\end{align}
This upper bound for $f= 1 {\rm TeV}$ is shown by the violet dashed line in Fig.~\ref{fig:mass range}. We find that for $T \agt 10^5 {\rm GeV}$, there is a parameter range, highlighted by light blue, where both Eqs.~(\ref{Eq:sigma}) and (\ref{constraint on mass range imposed by R.E.D.}) can be fulfilled. Notice that since $m/H$ is extremely large in this parameter range, realizing the onset of the oscillation in this parameter range requires a huge tuning of the potential gradient and the initial condition.

\section{Present day magnetic field strength}\label{Sec:present}
In this paper, we have discussed the generation of the seed magnetic field by an axion or an axion-like field which has commenced oscillation in a different epoch along the history of the Universe. We have also shown that the generated magnetic field can have non-zero helicity density. Now, let us estimate the possible maximum amplitude of the magnetic field and the coherent length predicted in the setups discussed in Sec.~\ref{Sec:gauge field generation during inflation}, \ref{Sec:production during axion reheating}, \ref{Sec:spectator production}, and \ref{Sec:post inflationary gen}. 

When the magnetic field evolves adiabatically, the current (comoving) strength and coherent scale of the magnetic field are related to those 
at the generated epoch, $a_{\rm gen}$, as
\begin{equation}
    B_{k_m, 0} = \left( \frac{a_{{\rm gen}}}{a_0} \right)^2 B_{k_{m, {\rm gen}}}\,, \qquad \lambda_{m, 0} = \left(\frac{a_0}{a_{{\rm gen}}}\right) \left(\frac{a_{{\rm gen}}}{k_{m,\, {\rm gen}}/(2\pi)}\right)\,.  \label{Exp:magNH}
\end{equation}
However, the magnetic field does not always evolve adiabatically especially in the early Universe filled with cosmological plasma. The magnetic field induces the velocity fields of the plasma. When the time scale of the velocity fields becomes smaller than the cosmological time scale, the turbulent motion arises in the plasma and the backreaction from the turbulent motion to the magnetic field can not be neglected. Due to the turbulent motion, the magnetic field energy on the turbulent scales is transferred to smaller scales (direct cascade). The resultant magnetic field evolution is different, depending on whether the magnetic field is helical or not.

Even if the gauge field is generated through the Chern-Simons coupling, it is not necessarily highly helical. In our scenario, the helicity density $\langle \mathfrak{H} \rangle$ was generated during the tachyonic growth before the onset of the oscillation. 
Therefore, if the tachyonic growth phase does not last sufficiently long, the resultant $\langle \mathfrak{H} \rangle$ is negligible and the generated magnetic field is almost non-helical.
Furthermore, when the backreaction becomes significant, the interaction between the two polarization modes washes out the generated $\langle \mathfrak{H} \rangle$, except for large scales where the mixing of the two polarization modes is suppressed, as pointed out by lattice simulations conducted in Ref.~\cite{Adshead:2015pva}.
In this case, the seed magnetic field would turn out to be non-helical.

Currently, the observations of TeV $\gamma$-ray from distant blazars imply the existence of the magnetic field and have lead to the lower bound~\cite{Taylor:2011bn},
\begin{align}
\label{eq:const_gamma}
B >10^{-17} {\rm G} \, \times
\begin{cases}
 ({1 \rm Mpc}/\lambda_0)^{1/2}  \,, & \text{for\,\,} \lambda <{1 \rm Mpc} \,, \\ 
\quad 1 \,, & \text{for\,\,} \lambda >{1 \rm Mpc} .
\end{cases}
\end{align}
In this section, first we demonstrate that it might be difficult for non-helical~(or weakly helical) magnetic field generated in our mechanism, to survive and seed the magnetic field implied by $\gamma$ ray observations, 
in particular, from the view point of the coherent scale. 
Then, following Refs.~\cite{Martin:2007ue, Durrer:2013pga}, we show that non-zero helicity of the generated magnetic field can ease such difficulties and there exists a possibility for coherent scales to grow to cosmological scales with some extent of the magnetic field strength.

\subsection{Non-helical magnetic fields}

After the completion of reheating, the magnetic field drives the velocity field of the cosmological plasma through the Lorentz force.
When the velocity field evolves and the time scale of the fluid velocity
is shorter than the cosmological scale, i.e., $L/v < 1/H$ where $v$ is the typical velocity of the plasma on the length scale $L$, the turbulent motion arises on the scale $L$. The magnetic field energy on the turbulent scales is transferred to small scales (direct cascade) and eventually dissipates. When the turbulence is fully developed, the energies in the plasma motion and in the magnetic field are in equipartition.
In equipartition, the plasma velocity induced by the magnetic field during the turbulent regime is
given by the Alfv\'en velocity, $v_{\rm A} \simeq B/\sqrt{4 \pi \rho}$ with the plasma energy density $\rho$. Therefore, the condition,~$L/{v_{\rm A}} > 1/H$, imposed at a scale factor $a$ during the radiation dominated era can provide an upper bound on the magnetic fields as
\begin{align}
B_0 <  4.5 \times 10^{-13} {\rm G} \left(\frac{h}{0.7} \right)^2
\left(\frac{a_0/a}{10^5} \right) 
\left(\frac{\lambda_0}{1{\rm pc}} \right)\,.
\label{eq:B_condition}
\end{align}
Consequentially, a violation of this condition implies that the magnetic field on the coherent scale would decay via the direct cascade process.

In the magnetic field generation scenario, "magnetogenesis from axion~($=$ inflaton) during inflation", in Sec.~\ref{Sec:gauge field generation during inflation}, we obtain the strict upper bound on the amplitude of the magnetic field and the coherent length at present from Eqs.~(\ref{Cond:Binf}), (\ref{Exp:kminf}), and (\ref{Exp:magNH}) as
\begin{align}
    & B_{k, 0} \ll  10^{13}{\rm G}\,\,  e^{- 2 {\cal N}_{\rm gen}} \left( 10^5 {\rm GeV} \over T_R \right)^2\left(\frac{g_{*{\rm s}0}}{g_{*{\rm sR}}}\right)^{\frac{2}{3}}\left(r_{\rm sr, {\rm gen}}\over 0.1\right)^{3\over 4} \cr
    & \qquad \qquad \qquad  \times \left({\cal{P}}_{\zeta_{\rm sr, {\rm gen}}}\over 10^{-9}\right)^{1\over 2}\left(1\over {\alpha {\cal{D}}}\right)^{1\over 2}\left(f/\Mp \over{10^{-2}}\right)^{1\over 2} \frac{1}{(\Delta (\ln k)_{\rm gen})^{1/2}}\,, 
\end{align}
and 
\begin{align}
    & \lambda_0 = 10^{-27}{\rm pc}\left(\frac{T_{\rm R}}{10^5{\rm GeV}}\right) e^{{\cal N}_{\rm gen}} \left(\frac{g_{*{\rm s}0}}{g_{*{\rm sR}}}\right)^{\frac{1}{3}} \cr
    & \qquad \qquad \qquad \times \left(\frac{f}{10^{-2}\Mp}\right) 
    \left(\frac{1}{\alpha}\right)
    \left(\frac{0.1}{r_{\rm sr, {\rm gen}}}\right)
    \left(\frac{10^{-9}}{{\cal{P}}_{\zeta_{\rm sr, {\rm gen}}}}\right)^{\frac{1}{2}}\left(\frac{\varepsilon_{{\rm sr}, {\rm gen}}}{\varepsilon_{\rm gen}}\right)^{\frac{1}{2}}\,.
\end{align}
For a larger ${\cal N}_{\rm gen}$, we can obtain a larger coherent scale. However, the strength of the generated magnetic field on the corresponding scale is highly suppressed.

It is possible to generate strong magnetic field on small scales through this mechanism. However, the condition to evade the dissipation, \eqref{eq:B_condition}, restricts the possible amplitude of the magnetic field on such scales more tightly, making the generation of the magnetic field implied in the $\gamma$-ray observation, Eq.~\eqref{eq:const_gamma}, more difficult.
This difficulty arises also in the mechanisms addressed in Sec.~\ref{Sec:production during axion reheating}, and \ref{Sec:spectator production}. 
In the next subsection, we explore the possibility that this difficulty may be circumvented for the helical magnetic field, since the inverse cascade can transfer the seed magnetic field to a larger scale.

\subsection{Helical magnetic field}
Since the Chern-Simons coupling violates parity symmetry, the magnetic field which was generated through this coupling with the axion has a difference in the amplitude between the two polarization modes, which leads to a non-zero helicity density. Because of this non-zero helicity, the coherent scale can shift to large scales as the Universe evolves.

\subsubsection{Inverse Cascade}\label{SSSec:inverse cascade}

As we have explained, during the generation of the gauge field, the prominent growth of the helicity density takes place. When the Universe started to be dominated by charged plasma, the conductivity in the Universe becomes high and the ideal MHD situation can be approximately realized.
According to Maxwell's equations, the comoving magnetic helicity density~$\langle\mathfrak{H}\rangle$ is conserved in the highly conductive Universe.

As mentioned in the previous subsection, the turbulent velocity evolves through the Lorentz force, and the magnetic field on the coherent scale decays to small scales through the turbulence.
However, the helicity density should be conserved. To keep the helicity conservation with decreasing magnetic field energy, the inverse-cascade happens:~some of the magnetic field energy is transferred to large scales and as a result, the coherent scales below the Hubble scale, grow. 

Calculating the inverse cascade process, in general, requires numerical simulation, which is beyond the scope of this work. In this paper, as a simplistic analysis, we heuristically take into account the inverse cascade via the time conservation of helicity density~\cite{Martin:2007ue, Durrer:2013pga}. To be more explicit, we assume that the helicity density $\langle\mathfrak{H}\rangle$ is conserved in time after the generation of the magnetic field. Assuming that the helical magnetic field has a peak at $k = k_{m,\, {\rm gen}}$, the helicity density $\langle\mathfrak{H}\rangle$ at the generation is given by 
\begin{equation}
\label{helical density in terms of B field}
\langle\mathfrak{H}\rangle {\simeq} \Delta (\ln k)a_{\rm gen}^3\frac{a_{\rm gen}}{k_{\rm gen}}B^2_{k_{m, {\rm gen}}}\,.
\end{equation} 

When the turbulence fully develops, the plasma motion and the magnetic field are in equipartition and the velocity reaches the Alfv\'en velocity, $v\sim B/\sqrt{\rho}$.
The evolution of magnetic field due to the turbulence continues until the recombination epoch~\cite{Banerjee:2004df, Durrer:2013pga}. 
Identifying it with the turbulence scale,~$a/k \sim v /H$, we can estimate the peak scale of the magnetic field as
\begin{equation}
\frac{a_{\rm rec}}{k_{m, {\rm rec}}} \eqsim \frac{B_{k_{m,\rm rec}}}{\rho_{\rm rec}/\textit{M}_\text{pl}}\,. \label{Exp:Brec}
\end{equation}
Assuming that the magnetic field is adiabatically diluted after the recombination epoch, i.e., 
\begin{equation}
    B_{k_m, 0} = \left( \frac{a_{\rm rec}}{a_0} \right)^2 B_{k_{m, {\rm  rec}}}\,, \qquad \lambda_{m, 0} = \frac{a_0}{a_{\rm rec}} \frac{a_{\rm rec}}{k_{m, {\rm rec}}/(2\pi)}\,,
\end{equation}
we can rewrite Eq.~(\ref{Exp:Brec}) as~\cite{Banerjee:2004df}
\begin{equation}
\label{Strength and scale relation inverse cascade}
B_{k_m, 0} \simeq 10^{-8}{ \rm G}\times\left(\frac{\lambda_{m,0}}{\rm Mpc}\right)\,,
\end{equation}
where we have inserted the total energy density at the recombination as given by
\begin{equation}
\label{rec energy density}
\left(\frac{a^3_{\rm rec}\rho_{\rm rec}}{a_0^3\textit{M}_\text{pl}}\right)^{1/3} \simeq \left[ (1+z_{\rm rec})\frac{T_0^4}{\textit{M}_\text{pl}} \right]^{1/3}\,.
\end{equation}
Here, we have used, $ 1 {\rm GeV}^{-1} = 6.4 \times 10^{-39}\text{Mpc} $ and $ 1{\rm G} = 6.8 \times 10^{-20} {\rm GeV}^2 $. 

Equating the helicity density evaluated at the recombination epoch which is expressed only in terms of $\lambda_{m, {\rm rec}}$ by using Eq.~(\ref{Exp:Brec}) with Eq.~(\ref{helical density in terms of B field}), we obtain
\begin{equation}
\label{Exp:Lhel}
\begin{aligned}
\lambda_{m,0}
\simeq 3 \times 10^{5}\text{Mpc}\left(\frac{a_{\rm gen}}{a_0}                  \right)\left(\frac{\lambda_{m,{\rm gen}}}{\text{GeV}^{-1}}\right)^{\frac{1}{3}}\left[\frac{\Delta(\ln k)_{\rm gen}}{\Delta (\ln k)_{rec}}\right]^{\frac{1}{3}}\left(\frac{{B_{k_{m, {\rm gen}}}}}{\text{GeV}^2}\right)^{\frac{2}{3}}\,.
\end{aligned}
\end{equation}
Inserting Eq.~(\ref{Exp:Lhel}) into Eq.~(\ref{Strength and scale relation inverse cascade}), we also can obtain the (peak) amplitude of the magnetic field at present. Although one may find the scale factor dependence, $\lambda_{m, 0} \propto a_{\rm gen}/a_0$ is counter-intuitive, this is a consequence of helicity conservation, which leads to $\lambda_{m, 0} \propto a_{\rm gen}/a_{\rm rec} \propto (a_{\rm gen}/a_0) \times (T_{\gamma, {\rm rec}}/ T_{\gamma, 0})$. 

\subsubsection{Estimation of the helical magnetic field}
In this paper, we have discussed three mechanisms which generate the helical (hyper) magnetic field. Here, using Eqs.~(\ref{Strength and scale relation inverse cascade}) and (\ref{Exp:Lhel}), we estimate the possible amplitude of the magnetic field at present.

\begin{itemize}
    \item Magnetogenesis from axion~($=$ inflaton) during inflation
\end{itemize}
Inserting Eqs.~(\ref{Cond:Binf}) and (\ref{Exp:kminf}) into Eq.~(\ref{Exp:Lhel}),  we obtain
\begin{align}\label{eq:pres_phys_inf}
\lambda_{m,0} & \ll    
10^3 \text{Mpc} ~e^{- {\cal N}_{{\rm gen}}}  \left(\frac{a_{{\rm end}}}{a_{\rm R}}\right) \left(\frac{10^5 {\rm GeV}}{T_{{\rm R}}}\right)
\left(\frac{g_{*{\rm s} 0}}{g_{*{\rm s\, R}}}\right)^{\frac{1}{3}}\left(\frac{f/\Mp}{10^{-2}}\right)^{\frac{2}{3}} \left(\frac{\varepsilon_{\rm sr,{\rm gen}}}{\varepsilon_{\rm gen}}\right)^{\frac{1}{6}} \nonumber \\
& \qquad \qquad \qquad \times \left(\frac{1}{ \alpha^2 D}\right)^{\frac{1}{3}}\left(\frac{r_{{\rm sr},{\rm gen}}}{0.1}\right)^{\frac{1}{6}}\left(\frac{{\cal{P}}_{\zeta_{\rm sr, {\rm gen}}}}{10^{-9}}\right)^{\frac{1}{6}} \left(\frac{1}{\Delta (\ln k)_{\rm rec}}\right)^{\frac{1}{3}} \,,
\end{align}
where we used
\begin{equation}
    \frac{a_{{\rm gen}}}{a_0} = e^{- {\cal N}_{{\rm gen}}} \left(\frac{a_{{\rm end}}}{a_{\rm R}}\right) \left(\frac{T_{\gamma, 0}}{T_{\rm R}}\right)\left(\frac{g_{*{\rm s}0}}{g_{*{\rm s\, R}}}\right)^{\frac{1}{3}} \,. \label{Exp:rationa}
\end{equation}
Here, we add the subscript $\rm R$ to the quantities evaluated at the completion of the reheating and ${\cal N}_{{\rm gen}}$ denotes the $e$-folding between the generation and the end of inflation. Following standard notation, $g_{*s}$ is the effective relativistic degrees of freedom related to the entropy density and in this article, $g_{*s0}$ and $g_{*s{\rm R}}$ are its values at the present and reheating epochs, respectively. 

Here, we assumed that $\alpha (\Mp/f)$ is much larger than 1, which is typically required for a sizable gauge field production (see the discussion in the beginning of Sec.~\ref{Sec:gauge field generation during inflation}). In order not to disturb the inflationary expansion, the backreaction needs to be negligibly small since the axion is the inflaton in this case. 
Because of the scale factor dependence $\lambda_{m, 0} \propto a_{\rm gen}/a_0$, discussed in the previous section, a lower reheating temperature relaxes the upper bound on $\lambda_{m, 0}$. 

When we assume that the equation of state $w$ is time independent between the the end of inflation and reheating completion, we can express the reheating temperature $T_{\rm R}$ as
\begin{equation}
\label{Exp:TR}
    T_{\rm R} = \left( \frac{90}{g_{{\rm eff}, {\rm R}} \pi^2} \right)^{\frac{1}{4}} (\Mp H_{\rm gen})^{\frac{1}{2}} \left( a_{\rm end} \over a_{\rm R} \right)^{\frac{3(1+w)}{4}}  \,. 
 \end{equation}
 Here, $g_{\rm eff, R}$ is the effective relativistic degrees of freedom during the reheating epoch.
 Inserting this expression into Eq.~(\ref{eq:pres_phys_inf}), we obtain
\begin{align}\label{eq:pres_phys_inf2}
\lambda_{m,0} & \ll    
0.01 \text{pc}\,  e^{- {\cal N}_{{\rm gen}}}  \left(\frac{a_{{\rm end}}}{a_{\rm R}}\right)^{\frac{1- 3w}{4}} \left(\frac{g_{*{\rm s} 0}}{g_{*{\rm s\, R}}}\right)^{\frac{1}{3}} \left(\frac{f/\Mp}{10^{-2}}\right)^{\frac{2}{3}} \left(\frac{\varepsilon_{\rm sr,{\rm gen}}}{\varepsilon_{\rm gen}}\right)^{\frac{1}{6}} \nonumber \\
& \qquad \qquad \qquad \times \left(\frac{1}{ \alpha^2 D}\right)^{\frac{1}{3}}\left(\frac{0.1}{r_{{\rm sr},{\rm gen}}}\right)^{\frac{1}{12}}\left(\frac{10^{-9}}{{\cal{P}}_{\zeta_{\rm sr, {\rm gen}}}}\right)^{\frac{1}{12}} \left(\frac{1}{\Delta (\ln k)_{\rm rec}}\right)^{\frac{1}{3}} \,,
\end{align}
where we assumed $g_{{\rm eff}, {\rm R}} ={\cal O}(10^2)$. 

Inserting this expression into Eq.~(\ref{Strength and scale relation inverse cascade}), we also obtain the possible maximum value of $B_{k_m, 0}$. Notice that the coherent length $\lambda_{m, 0}$ becomes larger for a lower energy scale of inflation $H_{\rm gen}$, which was replaced with $\sqrt{r_{{\rm sr}, {\rm gen}} {\cal P}_{ \zeta, {\rm sr}, {\rm gen}} }$ by using Eqs.~(\ref{Exp:HMp})\footnote{\label{foot} We can schematically understand the reason why a lower energy scale of inflation and an instantaneous reheating lead to a larger $\lambda_{m, 0}$ and $B_{k_m, 0}$ as follows. Assuming that the helicity density $\langle\mathfrak{H}\rangle$ is conserved in time after the generation of the magnetic field, we obtain $\lambda_{m, {\rm rec}}$, which gives the present coherent length as $\lambda_{m, 0} = \frac{a_0}{a_{\rm rec}} \lambda_{m, {\rm rec}}$, as
\begin{align}
    \lambda_{m, {\rm rec}} \propto  \frac{a_{\rm gen}}{a_{\rm rec}} (\lambda_{m, {\rm gen}} B_{k_m, {\rm gen}}^2)^{\frac{1}{3}} \propto \frac{a_{\rm gen}}{a_{\rm rec}}  H_{\rm gen}^{\frac{1}{3}}\,, \label{eq:lambdam0}
\end{align}
where we assume $\lambda_{m, {\rm gen}} \propto 1/ H_{\rm gen}$ and $B_{k_m, {\rm gen}} \propto \sqrt{\rho_{\rm gen}} \propto H_{\rm gen}$. I.e., we assume that the seed of the magnetic field was generated at a physical wavenumber which is proportional to the Hubble scale, using (a fraction of) the energy density of our Universe. Then, using Eqs.~(\ref{Exp:rationa}) and (\ref{Exp:TR}), we obtain 
\begin{align}
   B_{k_m, 0} \propto \lambda_{m, 0} \propto \lambda_{m, {\rm rec}} \propto  e^{- {\cal N}_{{\rm gen}}}  \left(\frac{a_{{\rm end}}}{a_{\rm R}}\right)^{\frac{1- 3w}{4}} \left( \Mp \over H_{\rm gen} \right)^{\frac{1}{6}}\,. 
\end{align}
When the energy scale of the Universe at the generation is higher, the more significant dilution after the generation compensates the larger amplitude of $B_{k_m, {\rm gen}}$, making $B_{k_m, 0}$ smaller for a larger $H_{\rm gen}$. This is a robust prediction as far as the assumptions mentioned below Eq.~(\ref{eq:lambdam0}) are fulfilled.}. Meanwhile, for $w < 1/3$, $\lambda_{m, 0}$ becomes the largest for the instantaneous reheating with $a_{\rm R} \simeq a_{\rm gen}$.

\begin{itemize}
    \item Resonant magnetogenesis from axion~($=$ inflaton) during reheating
\end{itemize}
Inserting Eqs.~(\ref{Exp:kmreh}) and (\ref{Cond:Breh2}) into Eq.~(\ref{Exp:Lhel}),  we obtain
\begin{align}
 & \lambda_{m,0} \alt   10^4 {\rm Mpc} \left(\frac{10^5 {\rm GeV}}{T_{\rm{R}}}\right)\left(\frac{g_{*{\rm s}0}}{g_{*{\rm sR}}}\right)^{\frac{1}{3}} \frac{a_{\rm gen}}{a_{\rm R}}
\left( \frac{f/\Mp}{10^{-2}} \right)^{1/3}
 \cr
& \qquad \qquad \qquad \qquad \times \left(\frac{r_{{\rm sr} ,{\rm gen}}}{0.1}\right)^{\frac{1}{6}} \left(\frac{{\cal{P}}_{\zeta_{{\rm sr} ,{\rm gen}}}}{10^{-9}}\right)^{\frac{1}{6}} \frac{{\rm min}[\, 1,\, {\alpha^{-1/3}} \,]}{(\Delta (\ln k)_{\rm rec})^{1/3}}   \,,  \label{Cond:rehinf}   
\end{align}
where we used $\Delta(\ln k)_{\rm gen} = {\cal O}(\alpha)$. 

The inequality in Eq.~(\ref{Cond:rehinf}) is saturated, when the backreaction becomes comparable to the contribution of the axion. Then, the two circular polarization modes start to interact, leading to the wash-out of the generated helicity density~\cite{Adshead:2015pva}. Since having a non-zero helicity density is crucial for the inverse cascade, let us consider the case when the upper bound of Eq.~(\ref{Cond:rehinf}) is never saturated, keeping the backreaction suppressed.
Replacing $a_{\rm end}$ with $a_{\rm gen}$ in Eq.~(\ref{Exp:TR}) and plugging it into Eq.~(\ref{Cond:rehinf}),
we obtain
\begin{align}
&\lambda_{m,0} \alt   {0.1} {\rm pc} 
\left(\frac{g_{*{\rm s}0}}{g_{*{\rm sR}}}\right)^{\frac{1}{3}} \left( \frac{a_{\rm end}}{a_{\rm R}} \right)^{\frac{1-3w}{4}}
\left( \frac{f/\Mp}{10^{-2}} \right)^{\frac{1}{3}}  \nonumber \\
& \qquad \qquad \qquad  \times  \left(\frac{0.1}{r_{{\rm sr} ,{\rm gen}}}\right)^{\frac{1}{12}}\left(\frac{10^{-9}}{{\cal P}_{\zeta, {\rm sr} ,{\rm gen}}}\right)^{\frac{1}{12}}  \frac{{\rm min}[\, 1,\, {\alpha^{-\frac{1}{3}}} \,]}{(\Delta (\ln k)_{\rm rec})^{\frac{1}{3}}}  \,,  \label{Cond:rehinf_RBR2}
\end{align}
where we again assumed $g_{{\rm eff}, {\rm R}} ={\cal O}(10^2)$.

When the axion, which plays the role of the inflaton has a shallow potential such as the pure natural potential (\ref{Pure Natural Potential}), with $f/\Mp \ll 1$, the energy scale of inflation tends to be lower~\cite{Nomura:2017ehb} and the parametric resonance becomes more efficient due to the delayed onset of the oscillation. The second aspect may render the instantaneous reheating (in the cosmological time scale) with $a_{\rm end} \simeq a_{\rm reh}$ more feasible. A detailed investigation on this is left for a future study.

Figure \ref{Fg:rhoreh} presents several examples where the tachyonic instability and subsequent resonance instability generate the seed magnetic field with non-zero helicity density. In particular for $\alpha = 2.6$ and $f/\Mp = 0.1$ (the green curves in the left panel), we obtain $B_h(\eta_{\rm gen},\, k_{m, {\rm gen}}) = 2.1\times10^{28} {\rm GeV^2}$ and $\lambda_{m, {\rm gen}} = 8.9\times 10^{-12}{\rm GeV}^{-1}$. Inserting these values into Eq.~(\ref{Exp:Lhel}), we obtain $\lambda_0 \simeq  0.3~{\rm pc}$ and $B_{k_{m, 0}} \simeq 3\times 10^{-15} {\rm G}$, where we use \eqref{Exp:TR} with $a_{\rm end} = a_{R}$. Here, the mass of the axion is $6.1 \times 10^{11} {\rm GeV}$, corresponding to $H_{\rm gen} = 5.4 \times 10^{10} {\rm GeV}$. In this case, $\rho_{\rm E}/\rho_{\phi}$, $\rho_{\rm B}/\rho_{\phi}$, and $|(\alpha/f)\langle \mathbf{E}\cdot\mathbf{B}\rangle|/| V_{,\, \phi}|$ all remain below $\sim 0.02$. 

In Refs.~\cite{Fujita:2015iga, Adshead:2016iae}, the possible amplitude of the magnetic field generated during the reheating after the axion inflation (with the quadratic potential) was discussed. Notice that (one of) the upper bounds on $\lambda_{m, 0}$ and $B_{k{m, 0}}$ comes from the upper bound on the energy density which can be transferred to the magnetic field. This upper bound should apply generically, as far as we generate the magnetic field using (a part of) the energy density of the Universe at the generation, $\rho_{\rm gen} = 3 H_{\rm gen}^2 \Mp^2$. Therefore, while the potential form of the axion is different, when we choose the corresponding model parameters, our constraint, Eq.~(\ref{Cond:rehinf_RBR2}), reproduces the upper bound argued in Refs.~\cite{Fujita:2015iga, Adshead:2016iae}, leaving aside the additional factor $(f/\Mp)^{1/3}$. This suppression factor is the consequence for realizing an efficient energy transfer from the axion to the gauge field without assuming a large coupling, characterized by $H_{\rm osc}/m \sim f/\Mp$. Nevertheless, as is shown in Fig.~\ref{Fg:rhoreh}, even with $f/\Mp =0.1$, the efficient energy transfer is possible for $\alpha = {\cal O}(1)$.

\begin{itemize} 
    \item Resonant magnetogenesis from axion~($\neq$ inflaton) during inflation or reheating
\end{itemize}
Inserting Eqs.~\eqref{Exp:kmreh}, \eqref{Cond:Bspc} and \eqref{Exp:TR} into Eq.~(\ref{Exp:Lhel}), we obtain 
\begin{align}
    \label{Cond:rehinfiso}
    \lambda_{m,0} \alt 0.1 {\rm pc}\,
    e^{-{\cal N}_{\rm end}}
    \left(\frac{a_{\rm end}}{a_{\rm R}}\right)^{\frac{1-3w}{4}}
    \left(\frac{g_{*{\rm s}0}}{g_{*{\rm s R}}}\right)^{\frac{1}{3}}
    \left(\frac{0.1}{r_{{\rm sr}, {\rm gen}}}\right)^{\frac{1}{12}}\left(\frac{10^{-9}}{{\cal{P}}_{\zeta, {\rm sr} ,{\rm gen}}}\right)^{\frac{1}{12}}
    \nonumber \\
    \times
    \left(\frac{F_{\rm gen}}{10^{-3}}\right)^{\frac{1}{3}}
    \left(\frac{H_{\rm gen}}{10^{-3m}}\right)^{\frac{1}{3}}
    \frac{{\rm min}[\, 1,\, \alpha^{- \frac{1}{3}} \,]}{(\Delta (\ln k)_{\rm rec})^{\frac{1}{3}}} \,. 
\end{align}
Notice that we obtain the same result as Eq.~(\ref{Cond:rehinf}), except for the suppression factor $e^{- {\cal N}_{{\rm gen}}} F_{{\rm gen}}^{1/3}$ and dependence of $H_{\rm gen}/m$ on different parameters. If the seed magnetic field was generated during reheating, we should set ${\cal N}_{{\rm gen}}$ to 0. 

\begin{itemize}
    \item Resonant magnetogenesis from axion in a plasma filled Universe
\end{itemize}
As was discussed in Sec.~\ref{Sec:post inflationary gen}, the resonant generation of the magnetic field after the creation of the plasma particles is somewhat challenging. Nevertheless, here, assuming that such a model can be constructed, let us evaluate the possible value of $\lambda_{m, 0}$. Similarly to the resonant production of the gauge field through the spectator axion during inflation or reheating, the peak wavenumber at the generation is given by Eq.~(\ref{Exp:kmreh}) and the maximum amplitude of the (hyper) magnetic field is given by Eq.~(\ref{Cond:Bspc}). Inserting these expressions into Eq.~(\ref{Exp:Lhel}), we obtain 
\begin{align}
 &   \lambda_{m,0} \alt   0.01 {\rm pc}  \left( F_{{\rm gen}} \over 0.1 \right)^{\frac{1}{3}} \left(\frac{g_{*{\rm s}0}}{g_{*{\rm s R}}}\right)^{\frac{1}{3}}
\left( 10^5 {\rm GeV} \over m \right)^{\frac{1}{3}} \left( T_{\rm gen} \over 10^5 {\rm GeV}\right)^{\frac{1}{3}}  \frac{{\rm min}[\, 1,\, \alpha^{-\frac{1}{3}} \,]}{(\Delta (\ln k)_{\rm rec})^{\frac{1}{3}}}   \,,
\end{align}
which, in turn, gives the upper bound on $B_{k_m, 0}$ by using Eq.~(\ref{Strength and scale relation inverse cascade}). 

\section{Conclusion}\label{Sec:conclusion}
In this paper, we have investigated the generation of the seed magnetic field through the Chern-Simons coupling between the U(1) gauge field and an axion~(or an axion-like field) with various mass scales. The axion commences oscillation, depending on its mass, at a different epoch in the history of the Universe. In this paper, we have addressed axions which begin oscillation during inflation, reheating, and also the radiation dominated era after the thermalization of the Universe. We have considered both cases where the axion is the dominant and subdominant cosmological component at the onset of oscillation, including axions which can play the role of the inflaton and dark matter, respectively.

The coherently oscillating axion leads to an exponential growth of the gauge field through the parametric resonance. In this paper, we have shown that even if the size of the dimensionless coupling $\alpha$ is of ${\cal O}(1)$, the parametric resonance can lead to a large amplification of the gauge field, when the onset of the oscillation delays, taking $H_{\rm osc}/m \ll 1$. While the parametric resonance enhances the two polarization modes with similar growth rates, the helicity density $\langle {\mathfrak{H}} \rangle$, which was first generated during the tachyonic instability before the onset of the oscillation, keeps on growing during parametric resonance. In particular, when the axion is the inflaton, both the amplitude and the helicity density of the gauge field are enhanced for $f/\Mp \ll 1$, because both the tachyonic instability and the parametric resonance persist for a longer duration for a smaller $f/\Mp$. When the generated gauge field is helical, a lower scale inflation and a quicker completion of reheating enhance the amplitude of the magnetic field at present (see footnote \ref{foot}). Notice that a shallow scalar potential which can realize the slow-roll evolution for $f/\Mp \ll 1$ tends to predict a low-energy-scale inflation because of a small $\varepsilon$. Furthermore, the efficient parametric resonance may lead to an instantaneous reheating on cosmological time scales.

When the axion, which amplifies the gauge field, is subdominant, the generation can take place at various epochs in the history of the Universe. Nevertheless, in this case, the amplitude of the predicted magnetic field is suppressed by the energy fraction of the axion to the total components in the Universe at the generation, $F_{\rm gen}$. We have shown that when a subdominant axion field enters an oscillatory phase during the radiation dominated epoch,  the electrical conductivity inhibits the amplification of gauge field and the generation of magnetic field is limited unless the potential parameters and initial condition are significantly tuned.

When the backreaction of the gauge field production becomes important, the parametric resonance typically terminates. Furthermore, the generated helicity density is washed out through a non-linear coupling between the two polarization modes via inhomogeneous modes of the axion. Taking into account these aspects, in this paper, we focused on the generation of the gauge field, assuming that the backreaction is perturbatively suppressed. Under this assumption, we have provided the order estimation of the magnetic field at present generated by the axion which started oscillating at different epochs in the history of the Universe. We find that the resonant amplification during the reheating can be a promising candidate for the generation of ${{\cal{O}} (10^{-16})~{\rm Gauss}}$, required by the lower limits imposed by blazar observations. Our order estimation of the present helical magnetic field is crucially based on the dynamics of the inverse cascade. A more accurate estimation requires magneto-hydrodynamic simulations in a cosmological setup.

In our scenario, during the generation of the gauge field, the electric component is also enhanced. Likewise in the flat space, the presence of a strong electric field can induce a large conductivity through the Schwinger effect~\cite{Schwinger:1951nm}. This can potentially terminate the generation of the seed magnetic field~(see Refs.~\cite{Domcke:2018eki,Frob:2014zka, Kobayashi:2014zza, Sobol:2018djj, Banyeres:2018aax} for the Schwinger effect in an inflationary spacetime). A detailed analysis is left for elsewhere, but here let us emphasize that similarly to the situation discussed in Sec.~\ref{Sec:post inflationary gen}, even if $\sigma/H \agt 1$, as far as $\sigma < {\rm Re}[\mu] m$, the resonant production of the seed magnetic field can take place.

\section*{Acknowledgements}
We would like to acknowledge Tatsuo Kobayashi for contributions at the early stage and thank P.~Adshead, A.~Basu, K.~Kamada, K.~Lozanov, and Y.~Tada for helpful discussions. T.~P. and Y.~U would like to thank Yukawa Institute for Theoretical Physics at Kyoto University, where this work was completed during the YITP-T-19-02 on "Resonant instabilities in cosmology". H.~T. is, in part, supported by Grant-in-Aid for Scientific Research~No.17H01110.
Y.~U. is supported by JSPS Grant-in-Aid for Young Scientists (B) under Contract No.~16K17689, Grant-in-Aid for Scientific Research on
Innovative Areas under Contract Nos.~16H01095 and 18H04349, Grant-in-Aid for Scientific Research (B) under Contract No. 19H01894, and the Deutsche Forschungsgemeinschaft (DFG, German Research Foundation) - Project number 315477589 - TRR 211. Y.~U. is also supported in part by Building of Consortia for the Development of Human Resources in Science and Technology and Daiko Foundation.


\appendix

\section{Numerical evaluation of the growth rate for spectator axion}
As has been discussed in the main part of this paper, the parametric resonance caused by the homogeneous oscillating axion can lead to an efficient generation of the gauge field.  (In Ref.~\cite{Hertzberg:2018zte}, the difference between the parametric resonance by the homogeneous axion and the one by a spherically symmetric axion clump was highlighted.) In particular, when the cosmic expansion is no longer important at the onset of the oscillation, i.e., $H_{\rm osc} \ll m$, the equation of motion \eqref{Eq:Adl} takes the form of the Hill's equation with the periodic frequency. Then, the Floquet theorem states that a solution of the Hill's equation is given by 
\begin{align}
\label{Exp:solHill}
    {\cal A}_h(t,\, k) = e^{\mu_k mt} P_h(mt) + e^{- \mu_k mt} Q_h(mt)\,,
\end{align}
where $P_h$ and $Q_h$ are periodic functions. Here, $\mu_k$ denotes the Floquet exponent. The growth rate is given by the real part of the Floquet exponent.

\begin{figure}
	\centering
	\includegraphics[width=0.9\linewidth]{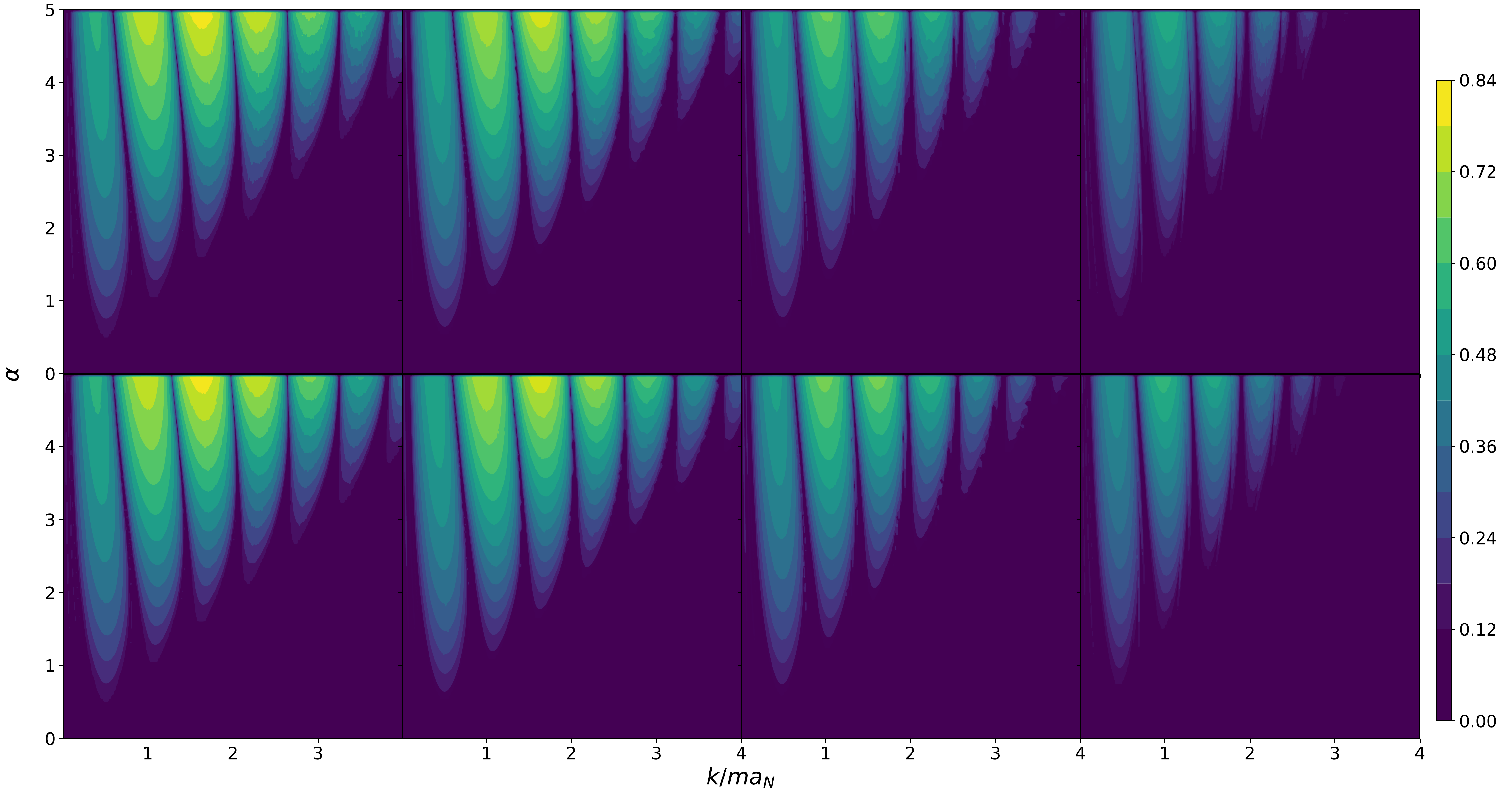}
	\caption{These two panels show the contours of the growth rate $\hat{\mu}_N$ during the $N=5$ oscillation cycle for $ H_{\text{osc}}/m = 0, 0.001, 0.005 \text{ and }0.01$ from left to right. Here, we consider the case where the axion is a spectator field. 
	The top and bottom panels represent the two circular polarization modes $h=-$ and $h=+$, respectively.}
	\label{fig:contour-of-growth-rate-with-different-hm}
\end{figure}

As was discussed in Sec.~\ref{Sec:spectator production}, when a spectator axion was initially located at a potential region whose gradient is shallower than the quadratic potential, the onset of the oscillation for the spectator axion delays, taking $H_{\rm osc}/m \ll 1$. In this appendix, to illustrate how the growth rate changes, depending on $\alpha$ and $H_{\rm osc}/m$, we consider a spectator axion which starts to oscillate during inflation. Here, instead of solving the time evolution of the spectator axion for a specific potential, we assume that $\tilde{\phi}$ evolves as $\tilde{\phi} = (a_{\rm osc}/a)^{3/2} \cos (mt)$ after starting the oscillation at various values of $H_{\rm osc}/m\,(\ll 1)$. Then, according to the Floquet theorem, the solution of ${\cal A}_h$ also becomes periodic over $m \Delta t = 2\pi$. For a computational brevity, instead of computing the Floquet exponent, here we simply calculate the growth rate during the $N$-th oscillation cycle defined as
\begin{equation}\label{Growth factor definition}
\hat{\mu}_N = \log\left[\frac{A_{\rm rms}(N)}{A_{\rm rms}(N-1)}\right]\Big/  (2\pi) \,,
\end{equation}
where $ A_{\rm rms}(N) $ denotes the root-mean-square value of $ |\sqrt{2k}\mathcal{A}_\pm|$ during the $N$-th oscillation cycle. When the decaying mode in Eq.~(\ref{Exp:solHill}) drops sufficiently fast, the growth rate defined in Eq.~(\ref{Growth factor definition}) provides a good approximation for the growth rate evaluated as ${\rm Re}[\mu]$. We take into account the cosmic expansion, using the scale factor $a = a_{\rm osc}\exp\left(H(t - t_{\rm osc})\right)$. 

Figure \ref{fig:contour-of-growth-rate-with-different-hm} shows the growth rate $\hat{\mu}_N$ for the 5-th oscillation cycle for the two circular polarization modes $h= \pm$. As is shown, both of $h= \pm$ undergo the exponentially growth. Since the interaction between the axion and the gauge field becomes weaker as the Universe expands, the growth late $\hat{\mu}_{N=5}$ becomes larger for a smaller value of $H_{\rm osc}/m$, for which the influence of the cosmic expansion is less significant. This can be also understood from the fact that the $q$-parameter in the Mathieu equation decreases as $q \propto \alpha (a_{\rm osc}/a)^{3/2}$. Notice that since we computed the growth rate given in Eq.~(\ref{Growth factor definition}) instead of the Floquet exponent, the decaying mode can contaminate the root-mean-square value $A_{\rm rms}(N)$, especially when the growth rate $\hat{\mu}_N$ is not large enough.    


\end{document}